\documentclass[10pt,aps,prd,floatfix,notitlepage,nofootinbib]{revtex4-1}

\usepackage{epsfig,float}
\usepackage{amsmath,amssymb,amsfonts}
\usepackage{multirow}
\usepackage{tikz}
\usetikzlibrary{matrix}

\usepackage{bm}

\begin{document}
\title{\bf Generalised teleparallel quintom dark energy non-minimally coupled  with the scalar torsion and a boundary term}




\author{Sebastian Bahamonde}
\email{sebastian.beltran.14@ucl.ac.uk}

\affiliation{Department of Mathematics, University College London, Gower Street, London, WC1E 6BT, United Kingdom}

\author{Mihai Marciu}
\email{mihai.marciu@drd.unibuc.ro}
\affiliation{Faculty of Physics, University of Bucharest, 405 Atomistilor, POB MG-11, RO-077125 Bucharest-Magurele, Roumania}

\author{Prabir Rudra}
\email{prudra.math@gmail.com}
\affiliation{Department of Mathematics, Asutosh College, Kolkata-700026, India.}

\date{\today}

\begin{abstract}
	Within this work, we propose a new generalised quintom dark energy model in
	the teleparallel alternative of general relativity theory, by considering a non-minimal coupling between the scalar fields of a quintom model with the scalar torsion component $T$  and the boundary term $B$. In the teleparallel alternative of general relativity theory, the boundary term represents the divergence of the torsion vector, $B=2\nabla_{\mu}T^{\mu}$, and is  related to the Ricci scalar $R$ and the torsion scalar $T$, by the fundamental relation: $R=-T+B$. We have investigated the dynamical properties of the present quintom scenario in the teleparallel alternative of general relativity theory by performing a dynamical system analysis in the case of decomposable exponential potentials. The study analysed the structure of the phase space, revealing the fundamental dynamical effects of the scalar torsion and boundary couplings in the case of a more general quintom scenario. Additionally, a numerical approach to the model is presented to analyse the cosmological evolution of the system.

\end{abstract}
\maketitle 

 \section{Introduction}
 
The appearance of Einstein's General Relativity in 1916 represents one of the most fundamental discoveries of modern physics. Since then, our understanding of the Universe has been improved, and new open problems have been revealed. Although various astrophysical experiments have been developed in order to study the general relativity theory, our cosmological understanding of the Universe is, so far, very limited. In the present cosmology, there are two fundamental questions which have triggered the main focus of theoretical and observational physics. One of them is represented by the dark matter fluid, a significant invisible component in the present Universe which acts gravitationally on large galactic scales. The other fundamental question of modern cosmology is represented by the the dark energy phenomenon. This phenomenon was discovered at the turn of the last century when two independent observational studies from distant Type Ia Supernovae (SNIa) revealed that the Universe is expanding at an accelerated rate \cite{exp1,exp2}. Considering the attractive nature of gravity, the existence of the dark energy phenomenon represented a paradigm shifting in the theoretical physics and cosmology. Since then, many theoretical speculations have been proposed in the scientific literature which can explain the dynamics and fundamental properties of this phenomenon \cite{de1}. The current astrophysical observations show that the Universe is made of approximately $27\%$ dark matter fluid, $68\%$ dark energy component, and less than $5\%$ usual baryonic matter \cite{compo,dm}. 
\par 
One of the most promising models for the dark energy phenomenon is represented by $\Lambda$CDM model \cite{lambda,de1}, where the dark energy is represented by a cosmological constant. In this model, the equation of state of dark energy is constant, having the value  $w=-1$. Although this model is favored by many cosmological observations, it has been shown that it suffers by various theoretical difficulties \cite{prob}. In order to solve the issues created by the $\Lambda$CDM model, the dynamical dark energy models have been proposed in the form of scalar field theories, where the dark energy equation of state is a dynamical component \cite{quint1,quint2,quint3}. In the case of quintessence models, the barotropic parameter of the dark energy equation of state is above the $\Lambda$CDM boundary, i.e., $w>-1$. For phantom type models, the dark energy is represented by a scalar field with a negative kinetic term \cite{phant1,phant2,phant3}, having a super-negative equation of state  $w<-1$, being slightly favored by recent observational data. 
\par 
Curiously, some of the present observational data are favoring a dynamical equation of state for the dark energy component corresponding to quintom models, with the dark energy equation of state parameter presenting an evolution from a phantom behavior $w<-1$ around present epoch, towards a quintessence behavior $w>-1$ in the near past \cite{super1,super2}. These observations have important astrophysical implications for the dark energy phenomenon: the cosmological constant has to be overruled due to the dynamical behavior of the dark energy equation of state; moreover, the quintessence and phantom models alone cannot explain the evolution of the dark energy equation of state and the possible crossing of the phantom divide line. Hence, a superposition between a phantom field and a quintessence field might explain the possible crossing of the phantom divide line by the dark energy component. The quintom dark energy model has emerged as a possible solution for the crossing of the phantom divide line by the dark energy component \cite{cross}. Although recent observations are consistent with the quintom scenarios \cite{nature1}, the Bayesian evidence still supports the $\Lambda$CDM cosmological model. In this regard, from the "no-go theorem", for the quintom scenarios \cite{quintobs}, we need to mention that a dynamically valid dark energy quintom model need to have at least two degrees of freedom. In the first proposed quintom model, the quintom fields were minimally coupled with gravity \cite{mot1}, and the action of the model simply consisted in the superposition between the phantom and the quintessence fields. Such a construction has been extended in the case of quintom scenarios with more general potentials in the work of Setare et. al.~\cite{mot2}. The case of mixed interactions between the quintom fields has been investigated by Saridakis et. al.~\cite{mot3}, by  considering the effects of a mixed term in the corresponding action of the model. 
\par 
In scalar tensor theories based on general relativity, the non-minimal couplings in the corresponding action have been seminally initiated in the work of Chernikov et. al.~\cite{nonminimal}. In these theories, it has been shown that the non-minimal couplings are an important aspect \cite{guage1,guage2,guage3,nonminimal2}, which might be taken into account in constructing a valid cosmological scenario. In scalar tensor theories, few papers have considered the effects of the non--minimal couplings in quintom constructions~\cite{citat1, citat2, citat3, citat4, citat5, citat6, citat7}. For the quintom paradigm, the non-minimal couplings have also been considered recently in the work of Marciu \cite{marciu1}, where the non-minimal couplings between the scalar quintom fields and the scalar curvature have been studied. This work showed that such a construction represents a feasible dark energy model, analyzing the effects of the non-minimal couplings for the dynamics of the Universe at large scale. 
\par 
In the recent years, in cosmological theories, a lot of attention has been focused towards the teleparallel equivalent of general relativity \cite{tegra,tegrb,tegr1,tegr2,tegr3}, representing an alternative theory to
general relativity which takes into account the torsion. In this theory, the torsion-less Levi-Civita connection is replaced by the curvature-less Weitzenb\"ock connection. It turns out that the teleparallel description gives rise to the standard Einstein field equations. Then, one can also modify the action to study some cosmological scenarios. 
One of the basic advantages of the teleparallel framework is represented by the fact that the torsion tensor consists exclusively in products of the first derivatives of the tetrad. Within this framework, an action can be formulated based on the torsion scalar $T$. In these theories, the torsion can be regarded as a counterpart of the curvature in scalar tensor theories based on general relativity. Hence, we have the following fundamental relation between the Ricci scalar and the torsion scalar \cite{baha}:
\begin{equation}
R=-T+B\,.
\end{equation} 
In this relation, $R$ is the Ricci Scalar, $T$ the scalar torsion, and $B$ a boundary term related
to the divergence of the torsion vector, being a total derivative. For more details about teleparallel theory, see \cite{pereira}.
\par 
In scalar tensor theories based on general relativity, an usual coupling function in the corresponding action is represented by a term like $\xi R\phi^2$, analysed by various authors \cite{nonminimal,nmc1,nmc2,Bamba:2013jqa}. An analogous formalism has been considered also in the teleparallel equivalent of general relativity, by taking into account non-minimal couplings of the type $\xi T\phi^2$ \cite{tegr1,tegr2,tel1,tel2,tel3,tel4,tel5,tel6}. Such constructions have revealed interesting consequences, like phantom behavior \cite{tegra,tegrb,tegr1,tegr2,tegr3} and the possible crossing of the phantom divide line for the dark energy equation of state. Hence, in the teleparallel equivalent of general relativity, interesting scenarios are represented in a dynamic model obtained by the non-minimal coupling between the quintessence field and a torsion scalar \cite{tegr1}. Recently, Bahamonde and Wright \cite{tegr3} extended this action \cite{tegr1} by adding a non-minimal coupling to the boundary term $B$, exploring the effects of the boundary coupling in the phase space structure, for an exponential potential. In the latter paper, the authors found that for the case where only a scalar field is coupled with $B$, the dynamic evolution of the system evolves towards a dark energy attractor at late times without fine tuning. They also showed that the phantom crossing barrier was also possible in this theory. The coupling of the teleparallel quintessence with the scalar torsion and the boundary term has been investigated for exponential and inverse power-law potentials, in the case of radiation and matter dominated epochs by considering scaling solutions of the Klein-Gordon equation \cite{marciu0}. Furthermore, the effects of the boundary couplings in the teleparallel alternative of general relativity have been investigated in various recent studies \cite{add1,add2,Zubair:2016uhx,add5, add6,add10,Bahamonde:2016cul,Bahamonde:2017wwk,Bahamonde:2016kba,Bahamonde:2017ifa,Bahamonde:2017sdo,sadjadi2017}. Let us emphasize here that when one is considering teleparallel theories of gravity or its modifications, there are two approaches in the literature: 1) One can work with the spin connection and find that the theory is local Lorentz invariant, however there is still a debate on how to compute the spin connection~\cite{Krssak:2015oua}; or 2) One can work in the pure  tetrad formalism setting the spin connection equal to zero, which gives a theory which is non covariant under Lorentz transformations, and then fixing the gauge by using the good tetrad approach \cite{Tamanini:2012hg}. Both approaches give the same equations since the spin connection does not change the equations. We will use the second approach since most of the literature in teleparallel gravity works with this approach. In principle, one can can switch on the spin connection and find a theory which is covariant under Lorentz transformations, but cosmologically speaking, the equations will be the same.  Recently, in \cite{Hohmann:2018vle}, it was presented a more general formalism for teleparallel scalar tensor theories which also considers a non-zero spin connection.  
\par 
In the case of quintom scenarios, Marciu~\cite{marciu1} proposed a quintom dark energy model with a non-minimal coupling between the scalar quintom fields and the scalar curvature in the theoretical framework of general relativity. Therefore, it is expected that such a construction can be extended to the teleparallel equivalent of general relativity. Motivated by \cite{tegr3} and \cite{marciu1}, we proceed to propose a more generalised teleparallel quintom dark energy model with non-minimal coupling to the scalar torsion and the boundary terms. Hence, the present paper can be regarded as an attempt aiming to unify the two previously approaches - quintom scalar tensor theories based on general relativity and quintom scalar tensor theories based on teleparallel gravity.
\par 
The paper is organised as follows. In section II, we propose the action of the generalised teleparallel quintom model with non-minimal couplings and we derive the corresponding evolution equations. In section III, we perform the dynamical system analysis of the present dark energy model and reveal the fundamental structure of the phase space. Then, in section IV, we perform a numerical approach by studying the evolution of the field equations for the present model and then analysing the implications of scalar torsion $T$ and boundary $B$ couplings for the dark energy equation of state. Finally, in section V we present the conclusions of our study and the final concluding remarks. The notation used in the paper is the same as in
\cite{tegr3}, where  tetrads and its inverse are denoted by a lower letter $e^{a}_{\mu}$ and a capital letter
$E^{\mu}_{a}$ respectively with the $(+,-,-,-)$ metric signature.

\section{Generalised Teleparallel quintom dark energy with non-minimal coupling}
In the same spirit as \cite{marciu1}, let us propose the following
generalised teleparallel action,
\begin{equation}
S = \int
\left[\frac{T}{2}+\frac{1}{2}\Big(f_1(\phi)+f_2(\sigma)\Big) T
+\frac{1}{2}\Big(g_1(\phi)+g_2(\sigma) \Big)B
+\frac{1}{2}\xi\partial_\mu \phi
\partial^\mu \phi+\frac{1}{2}\chi\partial_\mu \sigma\partial^\mu \sigma
-V(\phi,\sigma) +L_{\rm m}\right] e\, d^4x \,,\label{1}
\end{equation}
where $T$ is the scalar torsion, $B=2\nabla_{\mu}T^{\mu}$ the boundary term, $V(\phi,\sigma)$ is the potential
associated with both scalar fields and the functions $f_1,f_2,g_1$
and $g_2$ are coupling functions which depend on two different
scalar field $\phi$ and $\sigma$. Note that we have assumed $\kappa=1$. The constants $\xi$ and $\chi$
were introduced in order to have different kind of scalar fields.
If $\xi=\chi=1$ ($\xi=\chi=-1$) we will have two canonical scalar
field (two phantom scalar fields) whereas when $\xi=-\chi=1$
($-\xi=\chi=1$), the first scalar field $\phi$ will be canonical
and the second $\sigma$ will be phantom ($\phi$ phantom and
$\sigma$ canonical). Clearly, the action with one scalar field coupled non-minimally  with both the boundary term $B$ and the scalar torsion $T$ studied in
\cite{Zubair:2016uhx,tegr3} can be recovered if
\begin{eqnarray}
g_2(\sigma)=f_2(\sigma)=\chi=0\,,\quad V(\phi,\sigma)=V(\phi)\,,
\end{eqnarray}
so that, teleparallel dark energy (only one scalar field coupled to the scalar torsion $T$) can be recovered by choosing
\cite{tegr1}
\begin{eqnarray}
f_1(\phi)=c_1\phi^2\,,\quad
g_1(\phi)=g_2(\sigma)=f_2(\sigma)=\chi=0\,,\quad
V(\phi,\sigma)=V(\phi)\,.
\end{eqnarray}
Moreover, since $R=-T+B$, one can also recover the cases coming from the standard curvature framework. For instance, the quintom model non-minimally coupled with the scalar curvature $R$ studied in \cite{marciu1},
can be also recovered by setting
\begin{eqnarray}
f_1(\phi)=-g_1(\phi)=c_1\phi^2\,,\quad
f_2(\sigma)=-g_2(\sigma)=-c_1\sigma^2\,,\quad  -\xi=\chi=1\,.
\end{eqnarray}
Obviously, if one further assumes that $c_1=0$ and switches off one scalar field (for example setting $\sigma=0$), one could recover a scalar tensor theory non-minimally coupled with the scalar curvature $R$. Then, the above action represents a large class of teleparallel or
standard scalar tensor theories. By varying this action with
respect to the tetrads, we get
\begin{eqnarray*}
	2\left(1+f_1(\phi)+f_2(\sigma)\right)\left[e^{-1}\partial_\mu (e
	S_{a}{}^{\mu\nu})-E_{a}^{\lambda}T^{\rho}{}_{\mu\lambda}S_{\rho}{}^{\nu\mu}-\frac{1}{4}E^{\nu}_{a}T\right]-
	E^{\nu}_a \left[\frac{1}{2}\xi\partial_\mu \phi \partial^\mu
	\phi+\frac{1}{2} \chi\partial_\mu \sigma\partial^\mu \sigma
	-V(\phi,\sigma)\right]
\end{eqnarray*}
\begin{eqnarray*}
	+\xi E^{\mu}_a \partial^\nu \phi \partial_\mu \phi +\chi E^{\mu}_a
	\partial^\nu \sigma \partial_\mu \sigma+2\partial_{\mu}\Big(f_1(\phi)+f_2(\sigma)+g_1(\phi)+g_2(\sigma)\Big)
	E^\rho_a S_{\rho}{}^{\mu\nu}+E^{\nu}_{a}~^{\fbox{}}
	\left(g_1(\phi)+g_2(\sigma)\right)
\end{eqnarray*}
\begin{eqnarray}
-E^\mu_a
\nabla^{\nu}\nabla_{\mu}\left(g_1(\phi)+g_2(\sigma)\right)=\mathcal{T}^\nu_a\,,
\label{2}
\end{eqnarray}
where
$^{\fbox{}}={\nabla}_{\alpha}{\nabla}^{\alpha};~{\nabla}_{\alpha}$
is the covariant derivative linked with the Levi-Civita connection
symbol and $\mathcal{T}^{\nu}_a$ is the matter energy momentum tensor.\\

Now, variations of the action (\ref{1}) with respect to $\phi$ and
$\sigma$ gives us, respectively,
\begin{eqnarray}
\xi~^{\fbox{}} \phi+V_{\phi}&=&f_1'(\phi)T+g_1'(\phi)B\label{5a}\,,\\
\chi~^{\fbox{}}\sigma+V_{\sigma}&=&f_2'(\sigma)T+g_2'(\sigma)B\label{5b}\,,
\end{eqnarray}
where prime denotes differentiation with respect to the argument. Let us now study flat FRW cosmology where the metric is given by
\begin{equation}\label{6}
ds^2=dt^2-a^2(t)(dx^2+dy^2+dz^2)\ ,
\end{equation}
 with $a(t)$ the scale factor of the universe.
A good tetrads field \cite{Tamanini:2012hg} corresponding to this particular case is in the following form: $e^i_\mu=\textrm{diag}(1,a(t),a(t),a(t))$.
As usual, we will model the matter content of the universe
with a standard perfect fluid whose  energy-momentum tensor is defined as
\begin{equation}\label{7}
\mathcal{T}_{{\mu}{\nu}}=({\rho}_{\rm m}+p_{\rm m})u_{\mu}u_{\nu}-p_{\rm m}g_{{\mu}{\nu}}\,,
\end{equation}
where $u_{\mu}$ is the four velocity of the fluid and $\rho_{\rm m}$ and
$p_{\rm m}$ define the matter energy density and pressure, respectively.\\
For the FRW metric given by (\ref{6}), the field equations
(\ref{2}) become
\begin{eqnarray}
3H^{2}(1+f_1(\phi)+f_2(\sigma))&=&\rho_{\rm m}+V(\phi,\sigma)+\frac{1}{2}\xi\dot{\phi}^{2}+\frac{1}{2}\chi\dot{\sigma}^{2}
+3H(g'_1(\phi)\dot{\phi}+g'_2(\sigma)\dot{\sigma})\,,\label{FE1}\\
(3H^2+2\dot{H})(1+f_1(\phi)+f_2(\sigma))&=&-
p_{\rm m}+V(\phi,\sigma)-\frac{1}{2}\xi\dot{\phi}^{2}-\frac{1}{2}\chi\dot{\sigma}^{2}-2H(\dot{\phi}f_1'(\phi)
+\dot{\sigma}f_2'(\sigma))+\ddot{g_1}(\phi)+\ddot{g_2}(\sigma)\,,\label{FE2}
\end{eqnarray}
where $H=\dot{a}/a$ and dots represent differentiation with respect to
the cosmic time. The scalar fields equations
(\ref{5a}) and (\ref{5b}) become
\begin{eqnarray}
\xi(\ddot{\phi}+3H\dot{\phi})+3H^2f_1'(\phi)+3g_1'(\phi)\Big(3H^2+\dot{H}\Big)+\frac{V(\phi,\sigma)}{\partial \phi}&=&0\,,\label{12a}\\
\chi(\ddot{\sigma}+3H\dot{\sigma})+3H^2f_2'(\sigma)+3g_2'(\sigma)\Big(3H^2+\dot{H}\Big)+\frac{V(\phi,\sigma)}{\partial
	\sigma}&=&0\,,\label{12b}
\end{eqnarray}
where we have used $T=-6H^2$ and $B=-18H^2-6\dot{H}$. It can
be shown that the standard conservation equation for the fluids is
valid in this theory, i.e.,
\begin{eqnarray}\label{13}
\dot{\rho}_{\rm m}+3H(\rho_{\rm m}+p_{\rm m})=0\,.
\end{eqnarray}
Hence, if we assume a barotropic equation for the fluid
$p_{\rm m}=(\gamma-1) \rho_{\rm m}$, we directly find that
$\rho_{\rm m}=\rho_{0}a(t)^{-3\gamma}$, where $\gamma$ is a barotropic index and $\rho_0$ is an integration constant. $\gamma$ is physically
constrained to lie between $\gamma=0$, corresponding to a dark
fluid (behaving like a cosmological constant)  and $\gamma=2$,
corresponding to a stiff fluid. Therefore, we will only
concentrate our study in this range ($0\leq\gamma\leq 2$). For simplicity, let us assume that the energy
potential can be separated as follows
\begin{eqnarray}
V(\phi,\sigma)=V_1(\phi)+V_2(\sigma)\,.
\end{eqnarray}
The modified FRW equations (\ref{FE1}) and (\ref{FE2}) can be also be
rewritten in terms of effective energy and pressure, as follows
\begin{eqnarray}
3H^2&=&\rho_{\rm eff}\,,\\
3H^2+2\dot{H}&=&-p_{\rm eff}\,,
\end{eqnarray}
where we have defined $\rho_{\rm
	eff}=\rho_{\rm m}+\rho_{\phi}+\rho_{\sigma},~~ p_{\rm
	eff}=p_{\rm m}+p_{\phi}+p_{\sigma}$  and
\begin{eqnarray}
\rho_{\phi}=-3H^2f_{1}(\phi)+V_1(\phi)+\frac{1}{2}\xi \dot{\phi}^2+3Hg_1'(\phi)\dot{\phi}\,,\quad p_{\phi}=(3H^2+2\dot{H})f_{1}(\phi)+\frac{1}{2}\xi\dot{\phi}^2+2H\dot{\phi}f'_{1}(\phi)-\ddot{g}_{1}-V_{1}(\phi)\,,\\
\rho_{\sigma}=-3H^2f_{2}(\sigma)+V_2(\sigma)+\frac{1}{2}\chi
\dot{\sigma}^2+3Hg_2'(\sigma)\dot{\sigma}\,,\quad
p_{\sigma}=(3H^2+2\dot{H})f_{2}(\sigma)+\frac{1}{2}\chi\dot{\sigma}^2+2H\dot{\sigma}f'_{2}(\sigma)-\ddot{g}_{2}-V_{2}(\sigma)\,.
\end{eqnarray}

We can define the equation of state of the dark energy or scalar
fields as the following ratio of the scalar field pressures and
energy densities
\begin{equation}
\omega_{\phi}=\frac{p_{\phi}}{\rho_{\phi}}\,,~~~~~~~~~~\omega_{\sigma}=\frac{p_{\sigma}}{\rho_{\sigma}}\,.
\end{equation}
For the present generalised quintom model in teleparallel gravity, the dark energy equation of state is:
\begin{equation}
w_{\rm de}=\frac{p_{\phi}+p_{\sigma}}{\rho_{\phi}+\rho_{\sigma}}.
\end{equation}
We can also define the total or effective equation of state as
\begin{equation}
\omega_{\rm eff}=\frac{p_{\rm  eff}}{\rho_{\rm
		eff}}=\frac{p_{\rm m}+p_{\phi}+p_{\sigma}}{\rho_{\rm m}+\rho_{\phi}+\rho_{\sigma}}\,,
\end{equation}
and the standard matter energy density as:
\begin{equation}
\Omega_{\rm m}=\frac{\rho_{\rm m}}{3H^{2}}\,.
\end{equation}
Analogously, we define the energy density parameter for dark energy
or scalar fields as,
\begin{equation}
\Omega_{\phi}=\frac{\rho_{\phi}}{3H^{2}}\,,~~~~~~~~\Omega_{\sigma}=\frac{\rho_{\sigma}}{3H^{2}}\,,
\end{equation}
such that the relation
$\Omega_{\rm m}+\Omega_{\phi}+\Omega_{\sigma}=1$ holds.

\section{General Dynamical System}
In this section, we will study the dynamical system of the generalised teleparallel quintom model introduced in the last section. For a comprehensive description about dynamical systems in cosmology, see the review~\cite{baham1}. We consider the coupling constants as,
\begin{equation}
f_{1}(\phi)=c_{1}\phi^{2}\,,~~~f_{2}(\sigma)=c_{2}\sigma^{2}\,,~~~g_{1}(\phi)=c_{3}\phi^{2}\,,~~~g_{2}(\sigma)=c_{4}\sigma^{2}\,,
\end{equation}
where $c_i$ ($i=1,..,4$) are constants. Let us now introduce the dimensionless variables
\begin{equation}
s^2=\frac{\rho_{\rm m}}{3H^{2}},~~~
x^{2}=\frac{\dot{\phi}^{2}}{6H^{2}},~~~y^{2}=\frac{V_{1}(\phi)}{3H^{2}},~~~z=2\sqrt{6}
\xi
\phi,~~~u^{2}=\frac{\dot{\sigma}^{2}}{6H^{2}},~~~v^{2}=\frac{V_{2}(\sigma)}{3H^{2}},~~~w=2\sqrt{6}
\chi \sigma\,,
\end{equation}
which straightforwardly generalise the normalised variables used
to analyse standard quintessence \cite{Copeland}. If the Friedmann
equation \eqref{FE1} is written in terms of the above dimensionless
variables, it is reduced to the following constraint,
\begin{equation}
s^{2}=1-\xi
x^{2}-y^{2}+\frac{c_{1}}{24\xi^{2}}z^{2}-\frac{c_{3}}{\xi}xz-\chi
u^{2}-v^{2}+\frac{c_{2}}{24\chi^{2}}w^{2}-\frac{c_{4}}{\chi}uw\,.
\end{equation}
The above surface defines the boundary of our phase space.
Therefore, it is possible to reduce the dimensionality of the
dynamical system from 7 to 6.

We must also
assume that the energy density of matter is non-negative and since usually the potential is also considered to be positive in standard quintessence theories, we will further assume that condition. Hence, we must consider $y,v>0$.
$\xi,\chi,$ $\dot{\phi}$ while $\dot{\sigma}$ can be positive or negative. Thus, there is no restriction of signature
for $x,z,u$ and $w$.

For simplicity, we define the quantity $N = \log a$ and compute the dynamical equations in terms of its derivatives via $dx/dN=(1/H)dx/dt$.
In this variable, we obtain the following autonomous system of
first order
\begin{eqnarray}
\frac{dx}{dN}&=&\frac{1}{\sqrt{6}}\Big(q-\sqrt{6} p x\Big)\,,\label{xprime}\\
\frac{dy}{dN}&=&-\frac{y}{2}\Big(2 p +\sqrt{6} \lambda_{1} x \Big)\,,\label{yprime}\\
\frac{dz}{dN}&=&12 \xi  x\,,\label{zprime}\\
\frac{du}{dN}&=&\frac{1}{\sqrt{6}}\Big(q-\sqrt{6} p u\Big)\,,\label{uprime}\\
\frac{dv}{dN}&=&-\frac{v}{2}\Big(2 p + \sqrt{6} \lambda_{2} u \Big)\,,\label{vprime}\\
\frac{dw}{dN}&=&12 u \chi\label{wprime}\,,
\end{eqnarray}
where we have defined the following quantities
\begin{eqnarray}
p&=&-\frac{1}{2 \left(\chi ^3 \left(c_{1} \xi  z^2+6 c_{3}^2 z^2+24 \xi ^3\right)+c_{2} \xi ^3 w^2 \chi +6 c_{4}^2 \xi ^3 w^2\right)}\Big[3 \chi ^3 \Big(c_{1} z (4 c_{3} z+\xi  (16 \xi  x+\gamma  z))\nonumber\\
&&+4 \left(3 c_{3}^2 z^2-c_{3} \xi  \left(6 \xi  x (4 \xi  x+(\gamma -2) z)+\sqrt{6} \lambda_{1} y^2 z\right)-6 \xi ^3 \left(\gamma  \left(u^2 \chi +v^2+\xi  x^2+y^2-1\right)-2 \left(u^2 \chi +\xi  x^2\right)\right)\right)\Big)\nonumber\\
&&+c_{2} \xi ^3 w (4 c_{4} w+\chi  (16 u \chi +\gamma  w))+12 c_{4}^2 \xi ^3 w^2-4 c_{4} \xi ^3 \chi  \left(6 u \chi  (4 u \chi +(\gamma -2) w)+\sqrt{6} \lambda_{2} v^2 w\right)\Big]\,,\\
q&=&-\frac{\sqrt{6} c_{1} z+\sqrt{6} c_{3} (p+3) z+6 \xi  \left(\sqrt{6} \xi  x-\lambda_{1} y^2\right)}{2 \xi ^2}\,,\\
\lambda_{1}&=&-\frac{V_{1}'(\phi)}{
	V_{1}(\phi)}\,,\\
\lambda_{2}&=&-\frac{V_{2}'(\sigma)}{ V_{2}(\sigma)}\,.
\end{eqnarray}

The effective equation of state in terms of the dimensionless variables is
given by

\begin{eqnarray}
\omega_{\rm eff}&=&\frac{1}{\chi ^3 \left(z^2 \left(c_{1} \xi +6 c_{3}^2\right)+24 \xi ^3\right)+\xi ^3 w^2 \left(c_{2} \chi +6 c_{4}^2\right)}\Big[24 \gamma  \xi ^3 \chi ^3+4 c_{1} c_{3} \chi ^3 z^2+16 c_{1} \xi ^2 \chi ^3 x z+\gamma  c_{1} \xi  \chi ^3 z^2-c_{1} \xi  \chi ^3 z^2\nonumber\\
&&+8 \xi ^3 \chi ^2 u w (2 c_{2}-3 (\gamma -2) c_{4})+4 c_{2} c_{4} \xi ^3 w^2+\gamma  c_{2} \xi ^3 \chi  w^2-c_{2} \xi ^3 \chi  w^2+6 c_{3}^2 \chi ^3 z^2-96 c_{3} \xi ^3 \chi ^3 x^2-24 \gamma  c_{3} \xi ^2 \chi ^3 x z\nonumber\\
&&+48 c_{3} \xi ^2 \chi ^3 x z+4 \sqrt{6} c_{3} \xi  \chi ^3 \lambda_{1}y^2 z+6 c_{4}^2 \xi ^3 w^2-24 \xi ^3 \chi ^3 u^2 ((\gamma -2) \chi +4 c_{4})+4 \xi ^3 \chi  v^2 \left(\sqrt{6} c_{4} \lambda_{2} w-6 \gamma  \chi ^2\right)\nonumber\\
&&-24 \xi ^3 \chi ^3-24 \gamma  \xi ^4 \chi ^3 x^2+48 \xi ^4 \chi
^3 x^2-24 \gamma  \xi ^3 \chi ^3 y^2\Big]\,.
\end{eqnarray}
In the following, we will assume that the energy potential is
given by the following exponential type,
\begin{eqnarray}
V_{1}(\phi)=V_1 e^{-\lambda_1 \phi }\,,\quad V_{2}(\sigma)=V_2
e^{-\lambda_2 \phi }\,,\label{expon}
\end{eqnarray}
where $V_1,V_2$ and $\lambda_1,\lambda_2$ are constants. In our study, we will further assume that $\lambda_1>0$ and $\lambda_2>0$.

\subsection{The autonomous dynamical system} Let us denote the autonomous
system (\ref{xprime})-(\ref{wprime}) as
\begin{equation}
\dot{x}_{i}=f_{i}(x,y,z,u,v,w),~~~~~~~~~~~x_{i}=(x,y,z,u,v,w)\,.
\end{equation}
Critical or fixed points of Eqs.~(\ref{xprime})-(\ref{wprime})
correspond to $(x_{*}, y_{*}, z_{*}, u_{*}, v_{*}, w_{*})$ that
are solutions to all six equations $f_{i}(x_{*}, y_{*}, z_{*},
u_{*}, v_{*}, w_{*})=0$. There are 21 critical points for the
dynamical system, but only 13 satisfy $y\geq 0$ and $v\geq 0$
which ensures that the potentials are positive.  Table~\ref{Table1}
shows all the physical critical points of the system.
Table~\ref{Table2} presents the existence, acceleration criteria and stability regions for the points $O$, $A_{\pm}$ and $B_{\pm}$. For the remaining points, the expressions are complex and they will be presented in the forthcoming sections.
\begin{table}[H]
	\begin{center}
		\begin{tabular}{|c|c|c|c|c|c|c|}
			\hline
			Point & $x$ & $y$ & $z$ & $u$ & $v$ & $w$ \\
			\hline
			O & $0$ & $0$  & $0$ & $0$ & $0$ & $0$ \\
			\hline
			$A_{\pm}$ & $0$ & $0$ & $0$ & $0$ & $0$ & $\pm 2\sqrt{-\frac{6}{c_2}} \kappa  \chi$   \\
			\hline
			$B_{\pm}$ & $0$ & $0$ & $\pm 2\sqrt{-\frac{6}{c_1}} \kappa  \xi$ & $0$ & $0$ & $0$  \\
			\hline
			$C_{\pm}$ & $0$ & $0$ & $0$ & $0$ & $\frac{ \sqrt{2(c_{2}+3 c_{4})} \sqrt{\chi  (c_{2}+3 c_{4})\pm\Delta_{1}}}{\sqrt{c_{2}\chi}\lambda_{2} }$ & $\frac{2 \sqrt{6} (\chi  (c_{2}+3 c_{4})\pm\Delta_{1})}{c_{2}\lambda_{2}}$\\
			\hline
			$D_{\pm}$ & $0$ & $\frac{\sqrt{2(c_{1}+3 c_{3})} \sqrt{c_{1} \xi +3 c_{3} \xi \pm\Delta_{2}}}{\sqrt{c_{1}\xi}\lambda_{1} }$ & $\frac{2 \sqrt{6} (c_{1} \xi +3 c_{3} \xi \pm\Delta_{2})}{c_{1}\lambda_{1}}$& $0$ & $0$ & $0$  \\
			\hline
			$E_{\pm}$ & $0$ & $ \sqrt{\frac{c_{1}+3 c_{3}}{\sqrt{6}\lambda_1 \xi }z}$ & $z$ & $0$ & $\frac{\sqrt{c_{2}+3 c_{4}} \sqrt{12 c_{2} \lambda_{1} \xi ^2 \chi +36 c_{4} \lambda_{1} \xi ^2 \chi \pm\Delta_3 \sqrt{\lambda_{1}} \xi  \chi }}{\sqrt{6\chi c_2\lambda_1} \lambda_{2} \xi  }$ & $\frac{\chi  \left(12 \sqrt{\lambda_{1}} \xi  (c_{2}+3 c_{4})\pm\Delta_3\right)}{\sqrt{6\lambda_1} c_{2}  \lambda_{2} \xi }$  \\ \hline
			$F_{\pm}$ & $0$ & $ \sqrt{\frac{c_{1}+3 c_{3}}{\sqrt{6}\lambda_1 \xi }z}$ & $z$ & $0$ & $-\frac{\sqrt{c_{2}+3 c_{4}} \sqrt{12 c_{2} \lambda_{1} \xi ^2 \chi +36 c_{4} \lambda_{1} \xi ^2 \chi \pm\Delta_3 \sqrt{\lambda_{1}} \xi  \chi }}{\sqrt{6\chi c_2\lambda_1} \lambda_{2} \xi  }$ & $\frac{\chi  \left(12 \sqrt{\lambda_{1}} \xi  (c_{2}+3 c_{4})\pm\Delta_3\right)}{\sqrt{6\lambda_1} c_{2}  \lambda_{2} \xi }$ \\ \hline
		\end{tabular}
	\end{center}
	\caption{Critical points of the dynamical system (\ref{xprime})-(\ref{wprime}). For simplicity, we have defined the quantities $\Delta_1=\chi  \sqrt{(c_{2}+3 c_{4})^2-c_{2}\lambda_{2}^2}$, $\Delta_2=\xi  \sqrt{(c_{1}+3 c_{3})^2-c_{1} \lambda_{1}^2}$ and $\Delta_3=\sqrt{6} \sqrt{c_2 \lambda_2^2 z \left(4 \sqrt{6} \xi  (c_1+3 c_3)-c_1\lambda_1 z\right)+24 \lambda_1 \xi ^2 \left((c_2+3 c_4)^2-c_2 \lambda_{2}^2 \right)}$.}
	\label{Table1}
\end{table}
\begin{table}[H]
	\begin{center}
		\begin{tabular}{|c|c|c|c|}
			\hline
			Point & Existence & Acceleration & Stability \\ \hline
			O & Always & $\gamma<-1/3$ & Saddle point \\ \hline
			\multirow{2}{*}{    $A_{\pm}$} &    \multirow{2}{*}{$c_2<0$} &  \multirow{2}{*}{ $c_4>0\land c_2<-2c_4$} & Stable if $c_{4}>0\land c_{2}<-3 c_{4}\land c_{1}>\frac{c_{2} c_{3}}{c_{4}}\land \xi \geq \frac{24 c_{4} (c_{1} c_{4}-c_{2} c_{3})}{c_{2}^2}$\\
			& & & Unstable if $c_{4}<0\land c_{2}<0\land  c_{1}>\frac{c_{2} c_{3}}{c_{4}}\land \xi \geq \frac{24 c_{4} (c_{1} c_{4}-c_{2} c_{3})}{c_{2}^2}$\\ \hline
			\multirow{2}{*}{    $B_{\pm}$} &    \multirow{2}{*}{ $c_1<0$} & \multirow{2}{*}{ $c_3>0\land c_1<-2 c_3$} & Stable if $c_{3}>0\land c_{1}<-3 c_{3}\land c_{2}>\frac{c_{1} c_{4}}{c_{3}}\land \chi \geq \frac{24 c_{3} (c_{2} c_{3}-c_{1} c_{4})}{c_{1}^2}$ \\
			& & & Unstable if $c_{3}<0\land c_{1}<0\land  c_{2}>\frac{c_{1} c_{4}}{c_{3}}\land  \chi \geq \frac{24 c_{3} (c_{2} c_{3}-c_{1} c_{4})}{c_{1}^2}$\\ \hline
		\end{tabular}
	\end{center}
	\caption{Existence, acceleration condition and stability for the
		critical points $O$, $A_{\pm}$ and $B_{\pm}$} \label{Table2}
\end{table}
\par 
The point $O$ is always a saddle point and it represents a matter dominated era, as it usually appears in the dynamical analysis. The corresponding eigenvalues for the $O$ critical point are the following:
\begin{eqnarray}
\mu_{1}^{\rm O}&=&\frac{3 \gamma }{2}\,,\\
\mu_{\pm,2}^{\rm O}&=&\frac{3 (\gamma -2)}{4}\pm\frac{\sqrt{9 (\gamma -2) \left((\gamma -2) \chi +16 c_4\right)-96 c_2}}{4 \sqrt{\chi }}\,,\\
\mu_{\pm,3}^{\rm O}&=&\frac{3 (\gamma -2)}{4}\pm\frac{\sqrt{9 (\gamma -2) \left((\gamma -2) \xi +16 c_3\right)-96 c_1}}{4 \sqrt{\xi }}\,.
\end{eqnarray}
\par
\begin{figure}[H]
	\epsfig{file=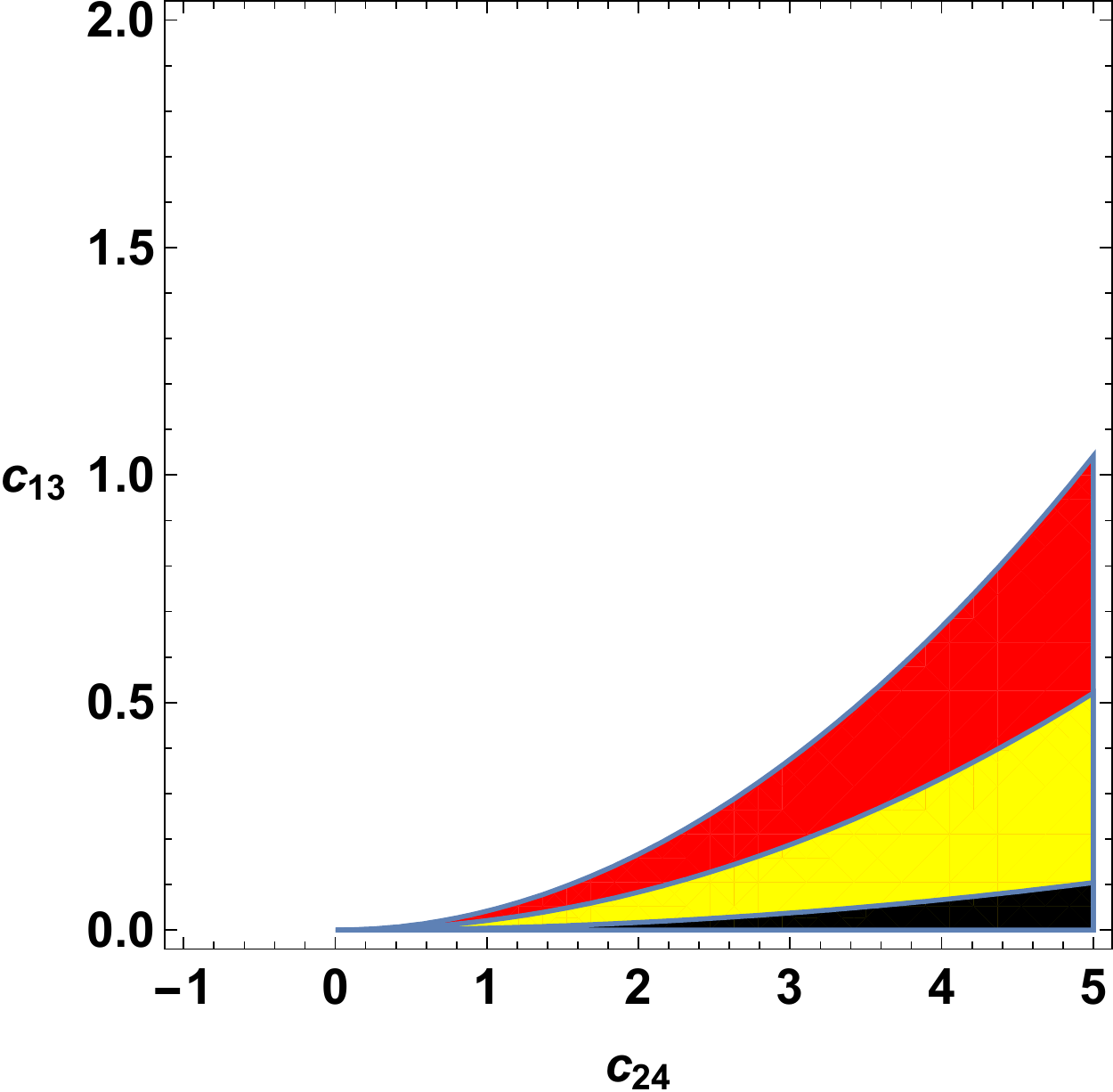, height=0.5\linewidth}
	\epsfig{file=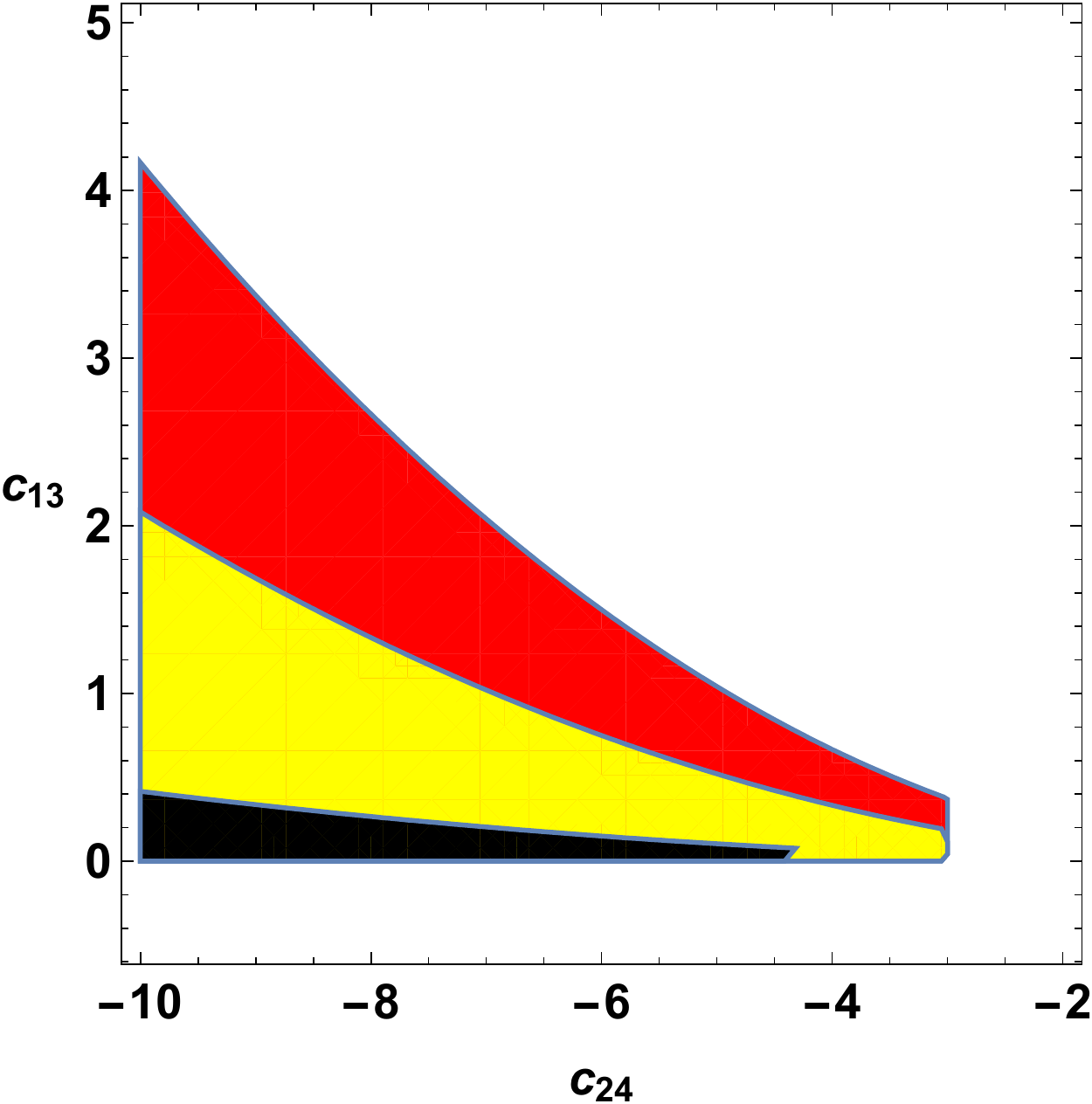,height=0.5\linewidth}
	\caption{The figure on the left(right) represents the regions where the points $A_{\pm}$ are unstable(stable). The black, yellow and red regions represent the cases where $\xi=0.1,0.5$ and $1$ respectively. The blank regions show the regions where the points are saddle points.
		We have defined the constants $c_{24}=c_2/c_4$ and $c_{13}=c_1-c_3c_{24}$ since the eigenvalues of those points
		only depend on those quantities. For the stability(instability) regions, we have assumed that $c_4>0$($c_4<0$) in order to have stability(unstability).}\label{Fig1}
\end{figure}

 
Critical points $A_+$ and $A_{-}$ correspond to a dynamical scenario where the first quintom field $\phi$ is absent, whereas the second quintom field $\sigma$ is frozen, without any kinetic or potential energy. The values of the $\sigma$ field are related to the $c_2$ parameter, describing the strength of the scalar torsion coupling for the $\sigma$ field. It can be noted from Table~\ref{Table2}, the points $A_{\pm}$ describe critical points corresponding to a dynamical scenario in which only the $\sigma$ field is non-negative, and the dark energy dominates the universe in terms of the density parameters. The eigenvalues of the $A_+$ and $A_{-}$ are the following:  
\begin{eqnarray}
\mu_{1}^{\rm A}&=&\frac{c_2}{c_4}+3\,,\\
\mu_{\pm,2}^{\rm A }&=&\frac{c_2}{2 c_4}\pm\frac{\sqrt{c_2^2 \xi +24 c_3 c_4 c_2-24 c_1 c_4^2}}{2 c_4 \sqrt{\xi }}\,,\\
\mu_{3}^{\rm A}&=&\frac{c_2}{c_4}\,,\\
\mu_{4}^{\rm A}&=&-3 \gamma +\frac{2 c_2}{c_4}+6\,.
\end{eqnarray}
Fig.~\ref{Fig1} shows the regions where the critical points $A_{\pm}$ are unstable and stable. Note that the point $A_+$ has the same eigenvalues as the point $A_-$. Thus, the stability regions of $A_-$ are the same as  $A_{+}$. It can be noticed that one needs negatives values of $c_{24}$ and positive values of $c_{13}$ in order to have stability. Remark that if $\xi$ is increased, the stability region also is increased.\\
Critical points $B_{\pm}$ correspond to a situation where the kinetic and the potential energy for the two quintom fields are equal to zero, the fields are frozen in time, and the first quintom field $\phi$ dominates the cosmic picture. The existence and acceleration conditions, as well as the stability criteria are expressed in Table~\ref{Table2}. The stability regions for the points $B_{\pm}$ have the same structure as Fig.~\ref{Fig1}, since those points are identical by changing $c_{1}\leftrightarrow c_{2},c_{4}\leftrightarrow c_{3}$ and $\xi\leftrightarrow \chi$. For the $A_{\pm}$ and $B_{\pm}$ critical points, we have to mention that the density parameter for the dark energy component is equal to $\Omega_{\rm de}=1$, describing a dark energy dominated universe.

\subsubsection{Critical points $C_{\pm}$ and $D_{\pm}$}
The following two classes of critical points, $C_{\pm}$ and $D_{\pm}$ are analysed in Table ~\ref{Table3}, where we show the existence conditions. For those two classes of critical points, the density parameters for the dark energy sector are equal to $\Omega_{\rm de}=1$, showing that the dark energy sector is dominating the cosmic picture. We should note that those four points are non-hyperbolic since  they have at least one zero eigenvalue. Since those points are non-hyperbolic, the conditions for stability cannot be determined using only standard linear stability theory, and another approach should be considered, such as center manifold method. Since our system is very complex and contains 6 independent parameters ($c_1, c_2, c_3, c_4, \lambda_1, \lambda_2$), such an approach cannot be applied symbolically due to the high complexity of the dynamical equations obtained after diagonalisation. Hence, in the present manuscript, we shall rely only on standard linear stability methods, and we shall analyse the eigenvalues for the corresponding critical points. Note again that if ones uses the linear stability method, it is not possible to determine the regions where the points are stable. However, if one eigenvalue is positive(negative) and at least another eigenvalue is negative(positive), then the point cannot be stable. Obviously, in principle there are more regions where the point can be saddle or unstable even if all the eigenvalues have the same sign. Those regions can be determined only by using other techniques, such as centre manifold or Lyapunov method. Thus, due to the difficulty of the system, we will only focus our study on determining the regions where the points are unstable or saddle by analysing the signs of the eigenvalues. 

\begin{figure}[H]
	\centering
\includegraphics[width=80mm]{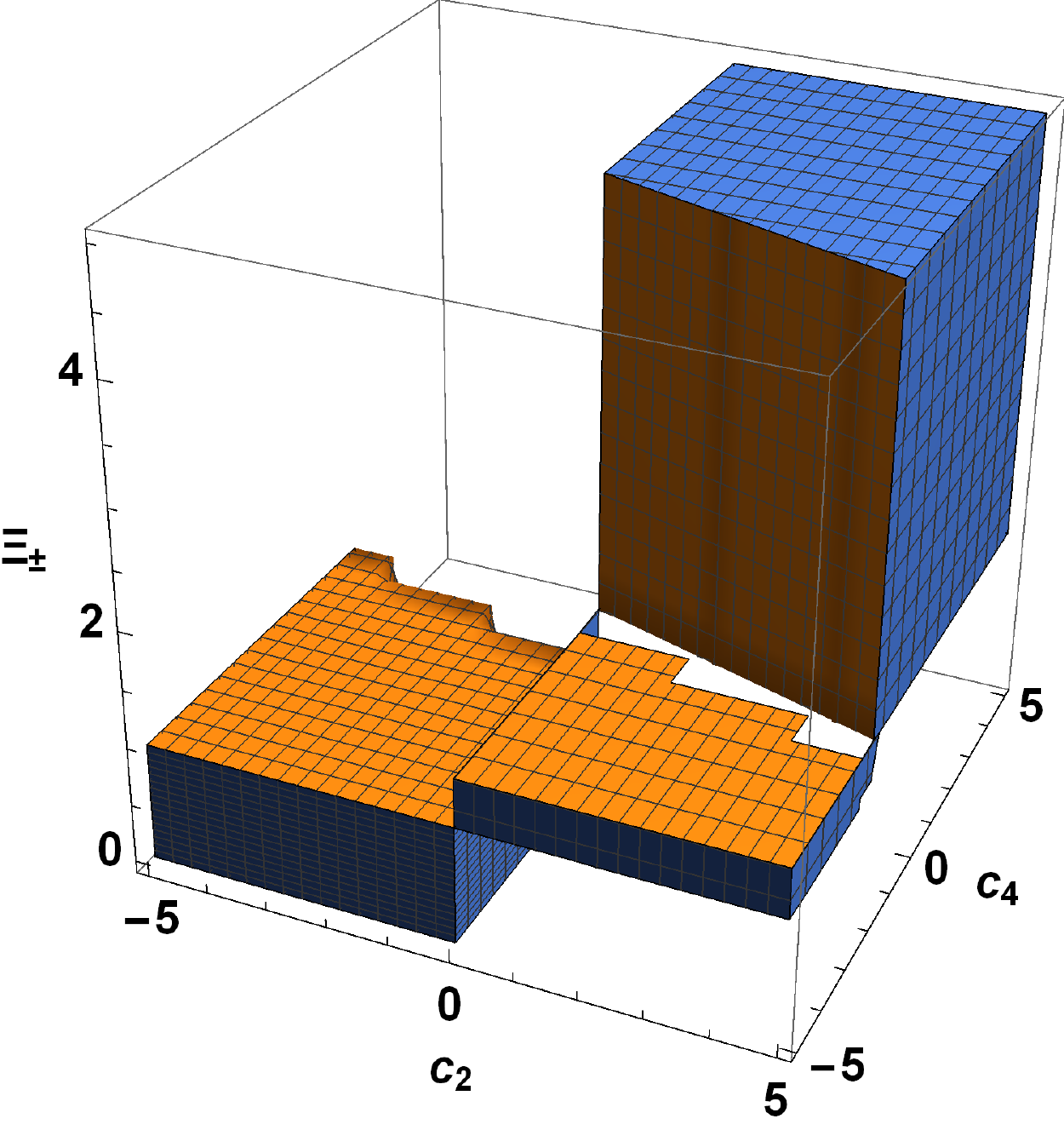}
\caption{Region plot for the points $C_\pm$ showing the regions where the points are always unstable or saddle. We have set $\xi=-\chi=-1$. Note that $\Xi_{+}$ corresponds to the point $C_{+}$, whereas $\Xi_{-}$ corresponds to the point $C_{-}$.}\label{FigCplus}
\end{figure}

In the following, we shall discuss the dynamical properties of those four points $C_{\pm}$, $D_{\pm}$, by considering the analysis of the corresponding specific eigenvalues. Analysing the Table ~\ref{Table1}, we can observe that points $C_{\pm}$ correspond to a scenario where the field $\phi$ is absent, having a zero kinetic and energy potentials. Also, for these critical points, the second field is frozen without any kinetic energy, having only a non-zero potential energy, while the value of the scalar function which describes the second field $\sigma$ is related to the parameters $c_2$ and $c_4$. The later constants describe the strength of the scalar torsion and boundary couplings for the $\sigma$ field.
	\begin{table}[H]
		\centering \begin{tabular}{|c|c|}
			\hline
			Point & Existence  \\ \hline
			\multirow{4}{*}{$C_{+}$} & $c_4\leq 0\land c_2<0\land \lambda_2\geq 0$\\
			& $c_4<0\land 0\leq \lambda_{2}\leq \frac{c_{2}+3 c_{4}}{\sqrt{c_{2}}}\land \Big(0<c_{2}\leq-3 c_{4}\lor c_{2}>-3 c_{4}\Big)$ \\
			& $c_{4}>0\land c_{2}\leq -3 c_{4}\land \lambda_{2}\geq 0$  \\
			&$c_4>0\land c_{2}>0\land 0\leq \lambda_{2}\leq \frac{c_{2}+3 c_{4}}{\sqrt{c_{2}}}$ \\ \hline
			\multirow{3}{*}{$C_{-}$} & $c_{4}<0\land 0\leq \lambda_{2}\leq \frac{c_{2}+3 c_{4}}{\sqrt{c_{2}}}\land \Big(0<c_{2}\leq-3 c_{4}\lor c_{2}>-3 c_{4}\Big)$  \\
			& $c_4\geq 0\land -3 c_{4}\leq c_{2}<0\land \lambda_{2}\geq 0$  \\
			& $c_4\geq 0\land c_{2}>0\land 0\leq \lambda_{2}\leq \frac{c_{2}+3 c_{4}}{\sqrt{c_{2}}}$  \\ \hline
			\multirow{4}{*}{$D_{+}$} & $c_3\leq 0\land c_1<0\land \lambda_1\geq 0$\\
			& $c_3<0\land 0\leq \lambda_{1}\leq \frac{c_{1}+3 c_{3}}{\sqrt{c_{1}}}\land \Big(0<c_{1}\leq-3 c_{3}\lor c_{1}>-3 c_{3}\Big)$ \\
			& $c_{3}>0\land c_{1}\leq -3 c_{3}\land \lambda_{1}\geq 0$  \\
			&$c_3>0\land c_{1}>0\land 0\leq \lambda_{1}\leq \frac{c_{1}+3 c_{3}}{\sqrt{c_{1}}}$  \\ \hline
			\multirow{3}{*}{$D_{-}$} & $c_{3}<0\land 0\leq \lambda_{1}\leq \frac{c_{1}+3 c_{3}}{\sqrt{c_{1}}}\land \Big(0<c_{1}\leq-3 c_{3}\lor c_{1}>-3 c_{3}\Big)$  \\
			& $c_3\geq 0\land -3 c_{3}\leq c_{1}<0\land \lambda_{1}\geq 0$ \\
			& $c_3\geq 0\land c_{1}>0\land 0\leq \lambda_{1}\leq \frac{c_{1}+3 c_{3}}{\sqrt{c_{1}}}$ \\ \hline
		\end{tabular}
		\caption{Existence conditions for points $C_{\pm}$ and $D_{\pm}$.}
		\label{Table3}
	\end{table}
In order to study the critical points $C_\pm$, it is convenient to introduce the following constants
\begin{equation}
\Xi_{\pm}=\frac{ \sqrt{2(c_2+3 c_4) \left(c_2+3 c_4\pm\sqrt{(c_2+3 c_4)^2-c_2 \lambda_{2}^2}\right)}}{\sqrt{c_2 \lambda_2}}\,.\label{Xi}
\end{equation}
Using the above constants, the eigenvalues of the points $C_\pm$ can be written as follows
\begin{eqnarray}
\lambda_{1}^{\rm C_\pm}&=&0\,,\label{Cplus1}\\
\lambda_{\pm,2}^{\rm C_\pm}&=&-\frac{3}{2}\pm\frac{\sqrt{-6 \left(8 c_2^2+24 c_2 c_4+9 c_4^2\right)+6 \Xi_{\pm}^2 (2 c_2+3 c_4)^2+9 c_2 \Xi_{\pm}^2 \chi }}{2 \sqrt{c_2 \Xi_{\pm}^2 \chi +6 c_4^2 \left(\Xi_{\pm}^2-1\right)}}\,,\\
\lambda_{\pm,3}^{\rm C_\pm}&=&-\frac{3}{2}\pm\frac{\sqrt{9 \xi -24 (c_1+3 c_3)}}{2 \sqrt{\xi }}\,,\\
\lambda_{4}^{\rm C_\pm}&=&-3\gamma\,.\label{Cplusf}
\end{eqnarray}
As we pointed out before, points $C_\pm$ are non-hyperbolic, since they have a zero eigenvalue ($\lambda_{1}^{\rm C_\pm}$). Since $\gamma\geq 0$, the eigenvalue $\lambda_{4}^{\rm C_\pm}<0$. Then, if one of the remaining eigenvalues ($\lambda_{\pm,2}^{\rm C_\pm}$ or $\lambda_{\pm,3}^{\rm C_\pm}$) are positive, the points $C_\pm$ cannot be stable, so they can be either saddle, or unstable. Further, eigenvalues  $\lambda_{-,2}^{\rm C_\pm}$ and $\lambda_{-,3}^{\rm C_\pm}$ are always negative. Then, they will not change the stability properties of the points. One can directly notice from the eigenvalue $\lambda_{+,3}^{\rm C_\pm}$ that the points $C_\pm$ are either saddle or unstable, if $\xi>0$ and $c_1<-3c_3$. Notice that for the existence of the points $C_+$($C_-$), we require $\Xi_+\geq 0$($\Xi_-\geq 0$) and $(\Xi_{+}-1)/c_2\geq 0$$(\Xi_{-}-1)/c_2\geq 0$). If $\chi >-(6 c_4^2 (\Xi_+^2-1))/(c_2\Xi_+^2)$, from $\lambda_{+,2}^{\rm C_+}$, one can also conclude that the point $C_+$ will be also saddle or unstable in all the following cases:
\begin{eqnarray}
c_4\leq 0\ \textrm{and}\ \left(\left(0<c_2<-3 c_4\ \textrm{and}\  1\leq \Xi_+<\sqrt{2}\right)\ \textrm{or}\  \left(c_2>-3 c_4\ \textrm{and}\  \Xi_+>\sqrt{2}\right)\right)\,,\\
\ \textrm{or}\  c_4>0,\ c_2>0 \ \textrm{and}\  \Xi_+>\sqrt{2}\,.
\end{eqnarray}
Additionally, for the case where $\chi <-(6 c_4^2 (\Xi_+^2-1))/(c_2\Xi_+^2)$, the point will be either saddle or unstable, when $c_4>0,\ -3 c_4<c_2<0$ and $0<\Xi_+\leq 1$.  The same conditions described above hold for $C_-$ by changing $\Xi_+\rightarrow \Xi_-$. Since the conditions written above are not so easy to understand, as an example, Fig.~\ref{FigCplus} shows a region plot for the specific case of quintom ($\xi=-\chi=-1$) for the points $C_\pm$. The regions represent the values of the parameters when the points are always saddle or unstable. For all the other cases, one needs to use other dynamical system techniques to conclude something about the stability of the points $C_\pm$.
\par
The following class of critical points presented in Table  ~\ref{Table3}, denoted as $D_{\pm}$, represents a similar behavior as the $C_{\pm}$ class, and shall be briefly discussed. We can notice that this class of critical points is related to a scenario in which the second quintom field $\sigma$ is absent, without any kinetic and potential energy, while for the first quintom field, $\phi$, we can notice that is is frozen, without any kinetic energy. However, the value of the field $\phi$ and the potential energy is related to the strengths of the $c_1$ and $c_3$ parameters, related to the scalar torsion and boundary couplings for the first quintom field, $\phi$. Since this class is similar to the previous class of critical points $C_{\pm}$, we shall omit it in the following presentation of the corresponding eigenvalues, as well as the stability figures, and we shall concentrate on the final classes of critical points. 
\subsubsection{Critical points $E_{\pm}$ and $F_{\pm}$}
\begin{figure}[H]
\centering
\includegraphics[width=.4\textwidth]{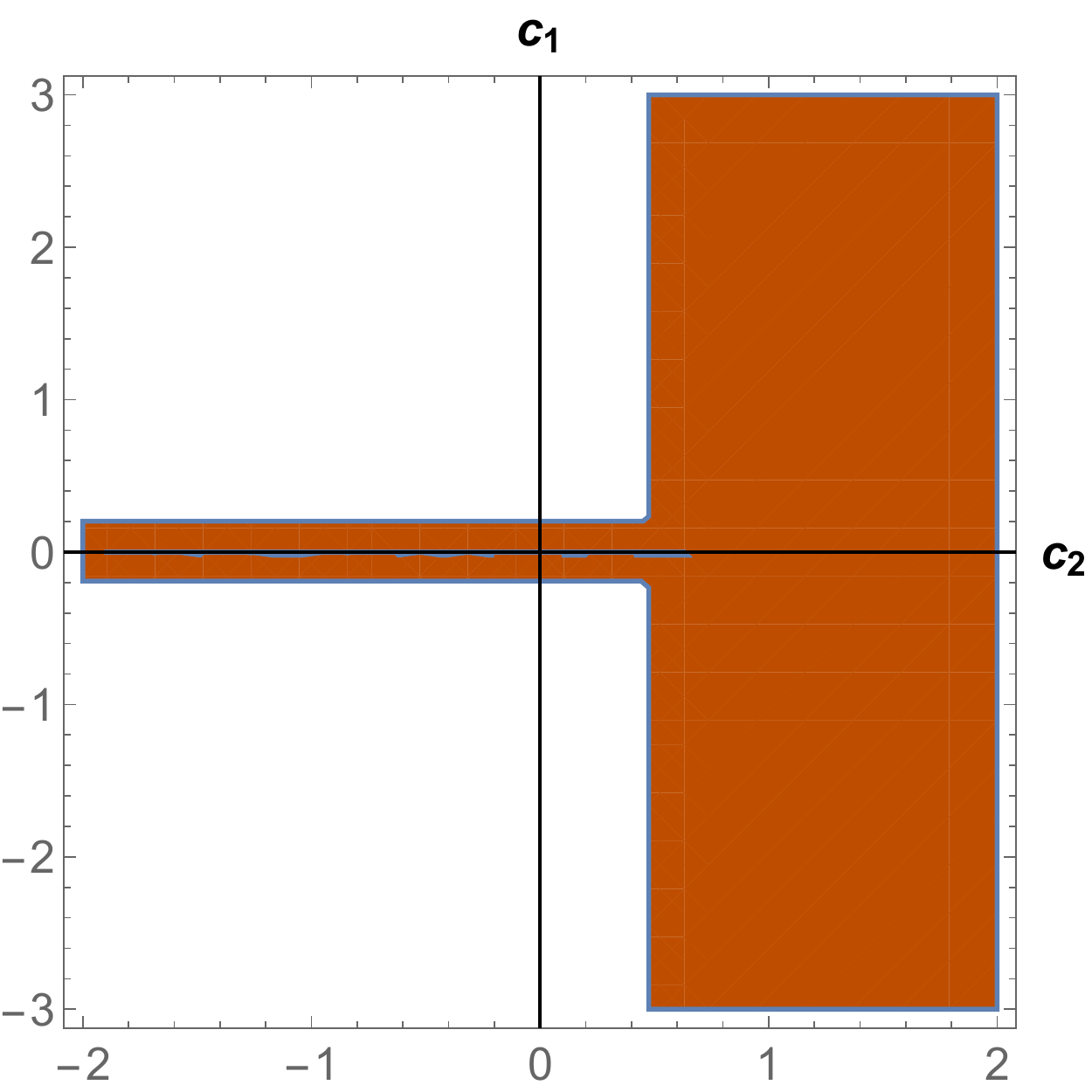}\quad
\includegraphics[width=.4\textwidth]{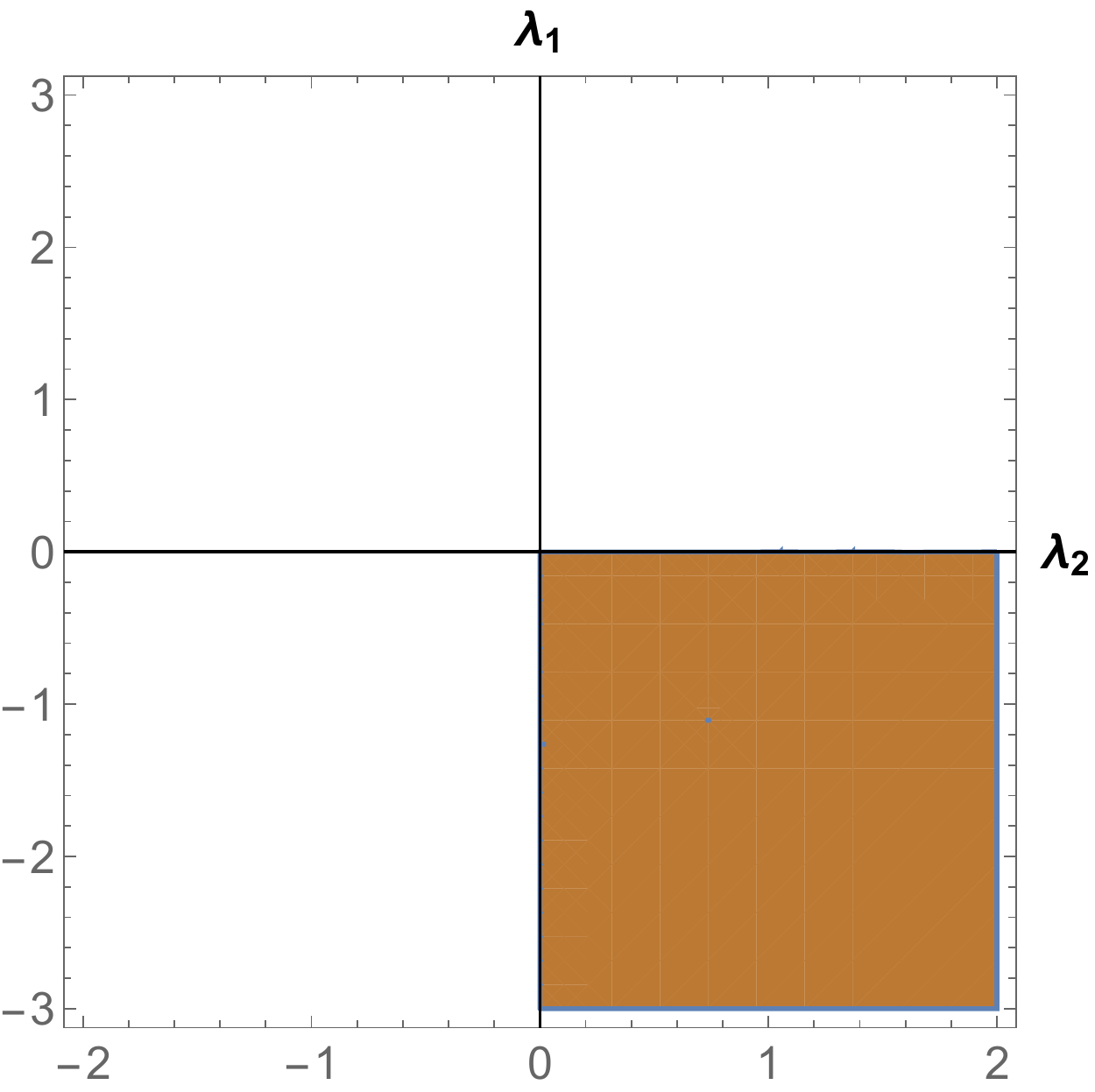}

\medskip

\caption{Region plots for which the point $E_{+}$ is in the real space and corresponds to a saddle or unstable behavior, where at least one eigenvalue has a positive real part and at least one eigenvalue has a negative real part. In the left panel, we have considered the following coefficients: $\xi=-1, \chi=+1, c_3 \approx -0.15, c_4 \approx 1.14, \lambda_1 \approx 1.73, \lambda_2 \approx -2.35$. In the right panel, we have considered the following values: $\xi=-1, \chi=+1, c_1=0.002, c_2=0.002, c_3 \approx -0.15, c_4 \approx 1.14$.}
\label{fig:figEplus}
\end{figure}

\begin{figure}[H]
\centering
\includegraphics[width=.4\textwidth]{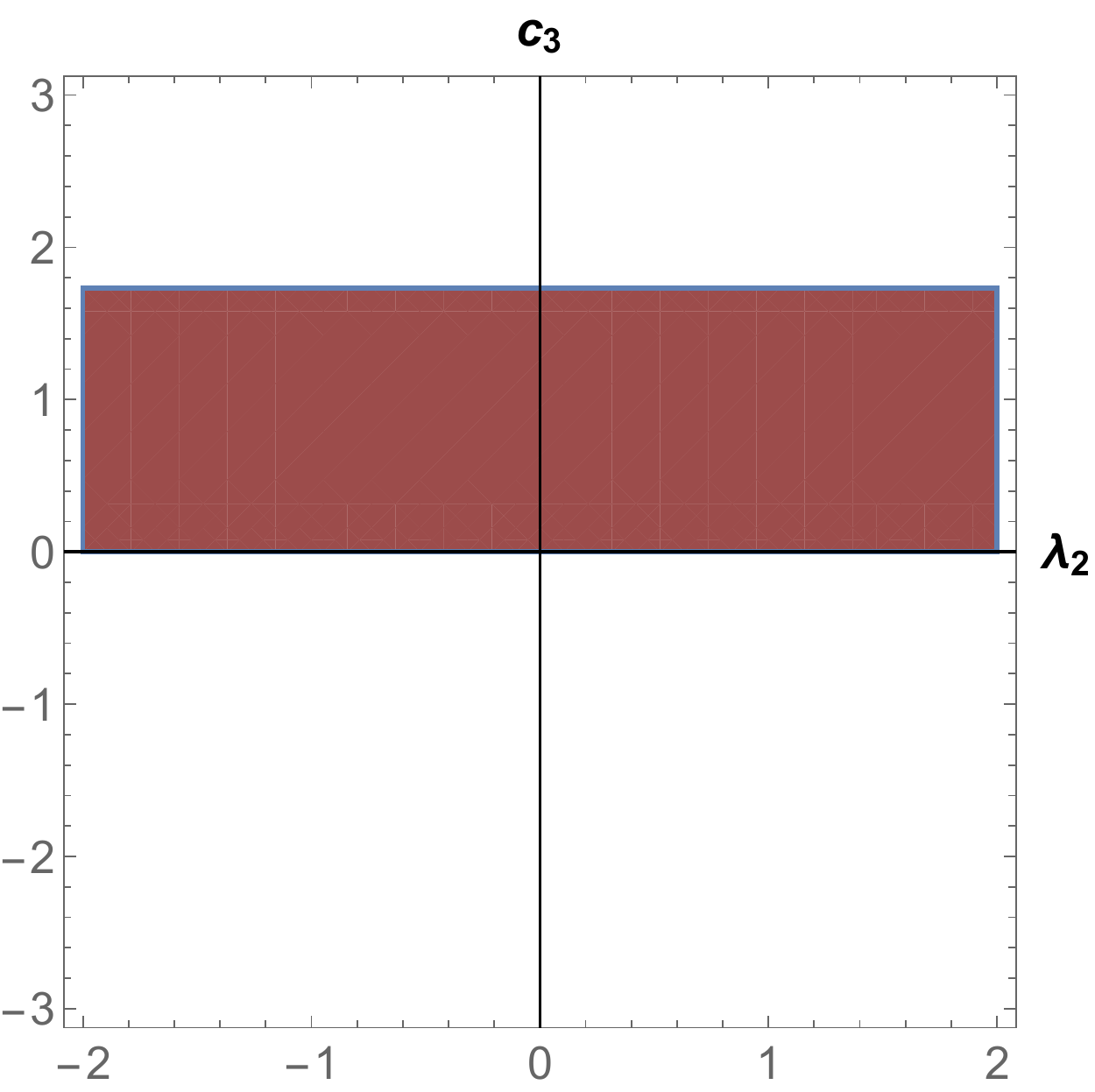}\quad
\includegraphics[width=.4\textwidth]{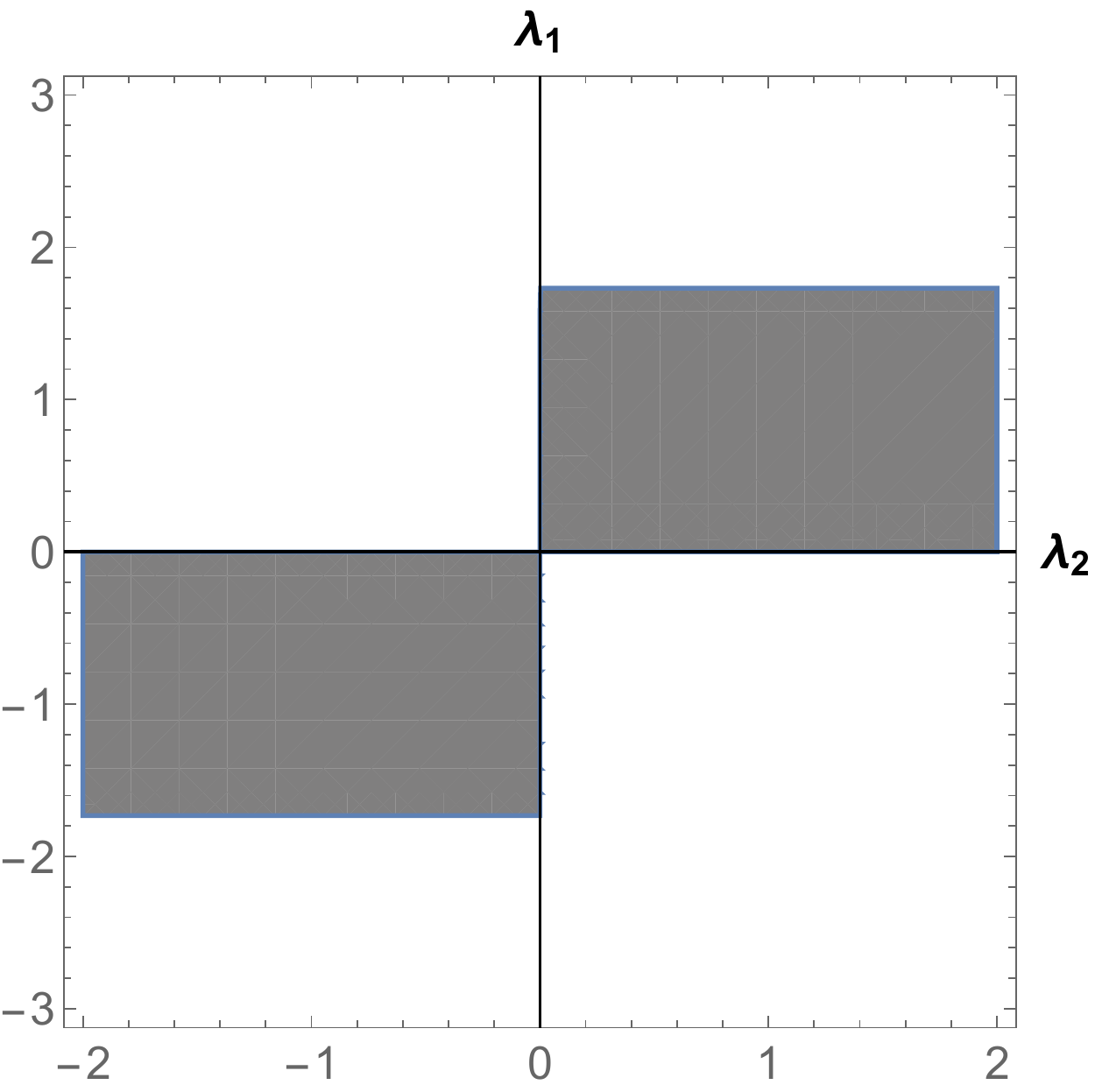}

\medskip

\caption{Domains for the case where the point $E_{-}$ is in the real space and corresponds to a saddle or unstable behavior. In the left panel, we have considered the following values of the coefficients for the present quintom model: $\xi=-1, \chi=+1, c_1=0.2, c_2=3, c_4=-2, \lambda_1=0.1$. For the right panel, we have the following values of the parameters: $\xi=-1, \chi=+1, c_1=0.2, c_2=3, c_3=2, c_4=-2$.}
\label{fig:figEminus}
\end{figure}

In the following, we shall discuss the dynamical properties of the last class of critical points, denoted in Table I as $E_{\pm}$ and $F_{\pm}$. As it can be observed from the table, those points are described by a critical line where the dynamical variable $z$ is independent, which is related to the scalar fiend $\phi$. For those critical lines, the dynamical variables $x$ and $u$, respectively are zero, which corresponds to a scenario in which the fields $\phi$ and $\sigma$ are frozen in time, without a kinetic energy for the two scalar fields $\phi$ and $\sigma$. In this case we can observe that for the $E_{\pm}$ and $F_{\pm}$ critical lines, we have a non-zero potential energy embedded into the $y$ and $v$ dynamical variables. The two classes of critical points, $E_{\pm}$ and $F_{\pm}$ are describing a universe which is characterised by the full domination of the dark energy fields over the matter component in the cosmos, $\Omega_{\rm m}=1-\Omega_{\rm de}=0$, and the dark energy equation of state is $w_{\rm de}=-1$, describing an accelerated scenario. Without assuming any value of $z$, the computation of the eigenvalues are very complicated.
In this case, we need to mention that for the $E_{\pm}$ and $F_{\pm}$ critical points, one eigenvalue is always zero. Then, we will perform a similar analysis as in the previous section, i.e., we will determine the regions where the points cannot be stable. Concerning the stability of those critical lines, in order to find the eigenvalues numerically, we will study the case where $z=0$. Hence, we shall concentrate on the specific case where the field $\phi$ is absent, displaying possible regions for the parameters of the model which correspond to saddle/unstable critical points where we have at least one eigenvalue with a positive real part and at least one eigenvalue with a negative real part (simultaneously). Since the dynamical properties of the critical points $E_{\pm}$ and $F_{\pm}$ are similar, we shall present our analysis only for the first class of critical points, $E_{+}$ and $E_{-}$. Considering the fine-tuning of the parameters of our model for the $E_{+}$ point, we have presented in Fig.~\ref{fig:figEplus} the case which corresponds to a saddle or unstable point, due to the presence to at least one positive eigenvalue and one negative eigenvalue, considering the cold dark matter equation of state for the matter component. In a similar way, for the $E_{-}$ critical point, we show some possible values of the parameters of the model which results in a saddle or unstable behavior for the $E_{-}$ point in Fig.~\ref{fig:figEminus}. The present discussion can be adapted also for the other critical lines in our dynamical system of equations, $F_{\pm}$.

\section{Numerical features of the quintom model}

\label{sec:trei}

\par
In this section we shall analyse the behavior of the present
generalised quintom model in the teleparallel gravity theory by
adopting a numerical approach previously considered by
Perivolaropoulos in scalar tensor theories
\cite{num1,num2}, and recently applied
to a quintom scenario \cite{marciu1}. The present
numerical approach enables us to analyse the effects of the
scalar torsion and boundary coupling in the evolution of the
corresponding quintom fields. In the following, we shall focus on
two distinct specific models, taking into account scalar torsion coupling
and boundary coupling, respectively, in an independent manner. We
shall investigate a quintom model, by taking the constants as:
$\xi=-\chi=+1$. At first, we shall analyse the following model
which takes into account only the scalar torsion coupling with the scalar fields:
\begin{equation}
f_1(\phi)=c_1\phi^2\,,\quad f_2(\sigma)=-c_1\sigma^2\,,\quad g_1(\phi)=0\,,\quad
g_2(\sigma)=0\,.
\end{equation}
\par
In the case of a decomposable scalar potential,
$V(\phi,\sigma)=V_1(\phi)+V_2(\sigma)$ given by eq.~\eqref{expon}, we have the following energy densities and pressures for the scalar quintom fields:
\begin{eqnarray}
\rho_{\phi}&=&\frac{1}{2}\dot{\phi}^2+V_1(\phi)-3c_1H^2\phi^2\,,\\
\rho_{\sigma}&=&-\frac{1}{2}\dot{\sigma}^2+V_2(\sigma)+3c_1H^2\sigma^2\,,\\
P_{\phi}&=&\frac{1}{2}\dot{\phi}^2+4c_1H\phi\dot{\phi}+c_1\phi^2(3H^2+2\dot{H})-V_1(\phi)\,,\\
P_{\sigma}&=&-\frac{1}{2}\dot{\sigma}^2-4c_1H\sigma\dot{\sigma}-c_1\sigma^2(3H^2+2\dot{H})-V_2(\sigma)\,.
\end{eqnarray}
\par
In this case, the Klein-Gordon equations which dictates the
behavior of the quintom fields become:
\begin{eqnarray}
\label{klein1}
\ddot{\phi}+3H\dot{\phi}+6c_1H^2\phi+\frac{dV_1(\phi)}{d\phi}&=&0\,,\\
\label{klein2}
\ddot{\sigma}+3H\dot{\sigma}+6c_1H^2\sigma-\frac{dV_2(\sigma)}{d\sigma}&=&0\,,
\end{eqnarray}
\par
with the dark energy equation of state being
\begin{equation}
w_{\rm de}=\frac{P_{\phi}+P_{\sigma}}{\rho_{\phi}+\rho_{\sigma}}=\frac{\frac{1}{2}\dot{\phi}^2+4c_1H\phi\dot{\phi}+c_1\phi^2(3H^2+2\dot{H})-V_1(\phi)-\frac{1}{2}\dot{\sigma}^2-4c_1H\sigma\dot{\sigma}-c_1\sigma^2(3H^2+2\dot{H})-V_2(\sigma)}{\frac{1}{2}\dot{\phi}^2+V_1(\phi)-3c_1H^2\phi^2-\frac{1}{2}\dot{\sigma}^2+V_2(\sigma)+3c_1H^2\sigma^2}\,.
\end{equation}
\par
In order to study the numerical evolution of the quintom model by
taking into account a scalar torsion coupling with the scalar fields, we need to express the evolution of the matter component in the
acceleration equation, namely \cite{num1,num2,marciu1}
\begin{equation}
\label{accelerationeqs}
\frac{\ddot{a}}{a}=-\frac{1}{6}(\rho_{\phi}+\rho_{\sigma}+3
P_{\phi}+ 3P_{\sigma} )-\frac{\Omega_{\rm m 0}H_{0}^2}{2 a^3}\,.
\end{equation}
\par
In this equation, $\Omega_{\rm m 0}$ is the density parameter for
the matter component at the present time, denoted as $t_0 \sim
0.96$. Moreover, $H_{0}$ and $a_{0}$ are the Hubble parameter and
cosmic scale factor at the present time. For further details on
the numerical implementation, the reader might consult
Refs.~\cite{num1,num2,marciu1}.
Further, we shall take the following values at the present time
$\Omega_{\rm m 0} \sim 0.30$ and $H_{0} \sim a_{0}\sim 1$, while at the
initial time we consider that the Universe is deep in the matter
epoch:
\begin{equation}
\label{eqqa} a(t) \sim  \Big(\frac{9 \Omega_{\rm m 0}}{4} \Big)^{1/3}
t^{\frac{2}{3}}\,.
\end{equation}
\par
Hence, our dynamical system of equations is now reduced to the two
Klein-Gordon equations \eqref{klein1} and \eqref{klein2}, together with
the acceleration equation \eqref{accelerationeqs}. We evolve
in time this reduced system of equations for different values of
the parameters of the model. The numerical solutions obtained  are
to be constraint by the requirement that at the present time
$t_0\sim 0.96$ we had the following relations: $\Omega_{\rm m 0}\sim
1- \Omega_{\rm de}\sim 0.30$, $H_{0} \sim a_{0}\sim 1$, where
$\Omega_{\rm de}$ represents the quintom energy density parameter,
deduced from the Friedmann constraint, with the
energy potential being considered as exponential type as Eq.~\eqref{expon}.

\begin{figure}[H]
\centering
\includegraphics[width=.4\textwidth]{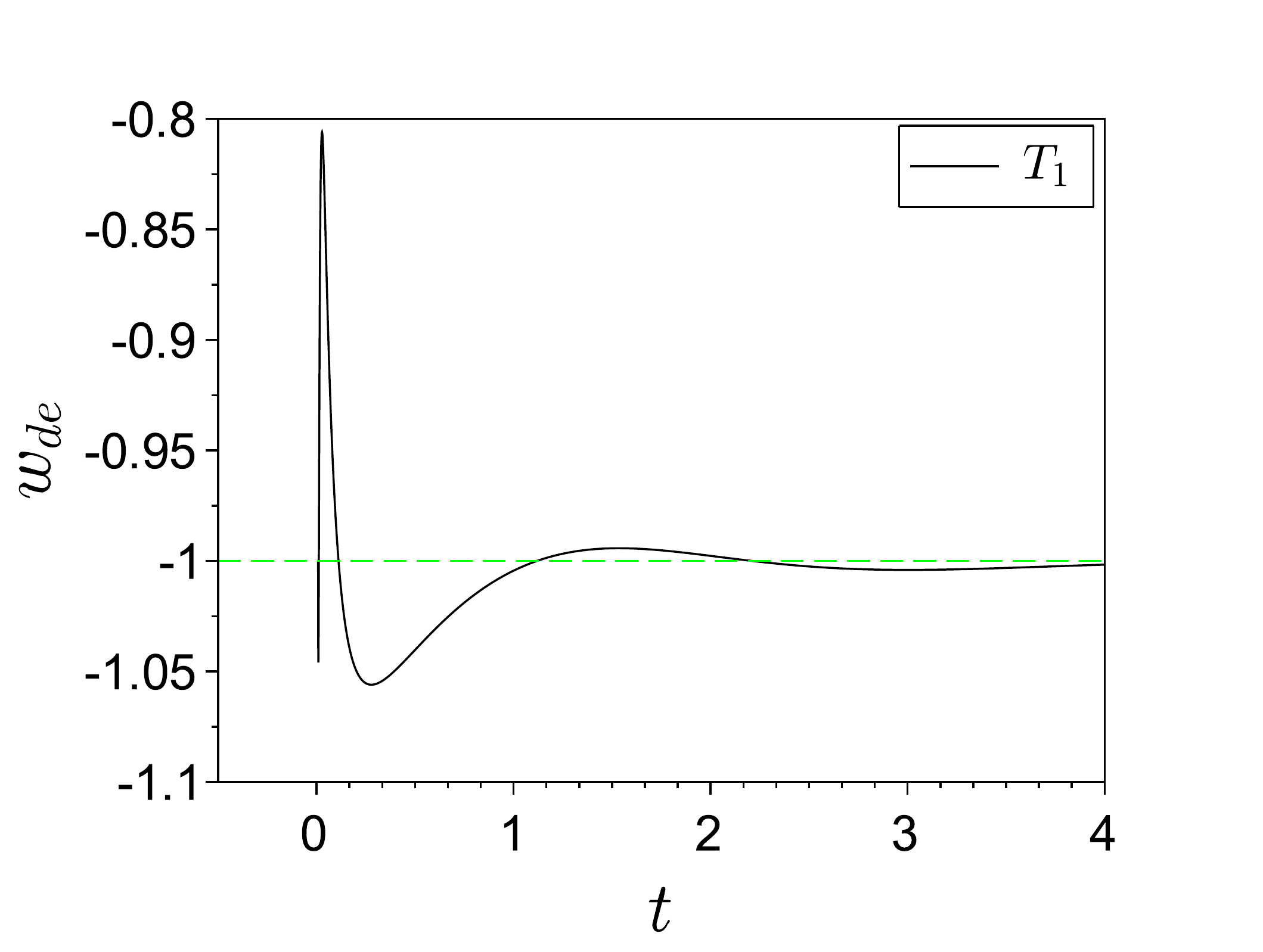}\quad
\includegraphics[width=.4\textwidth]{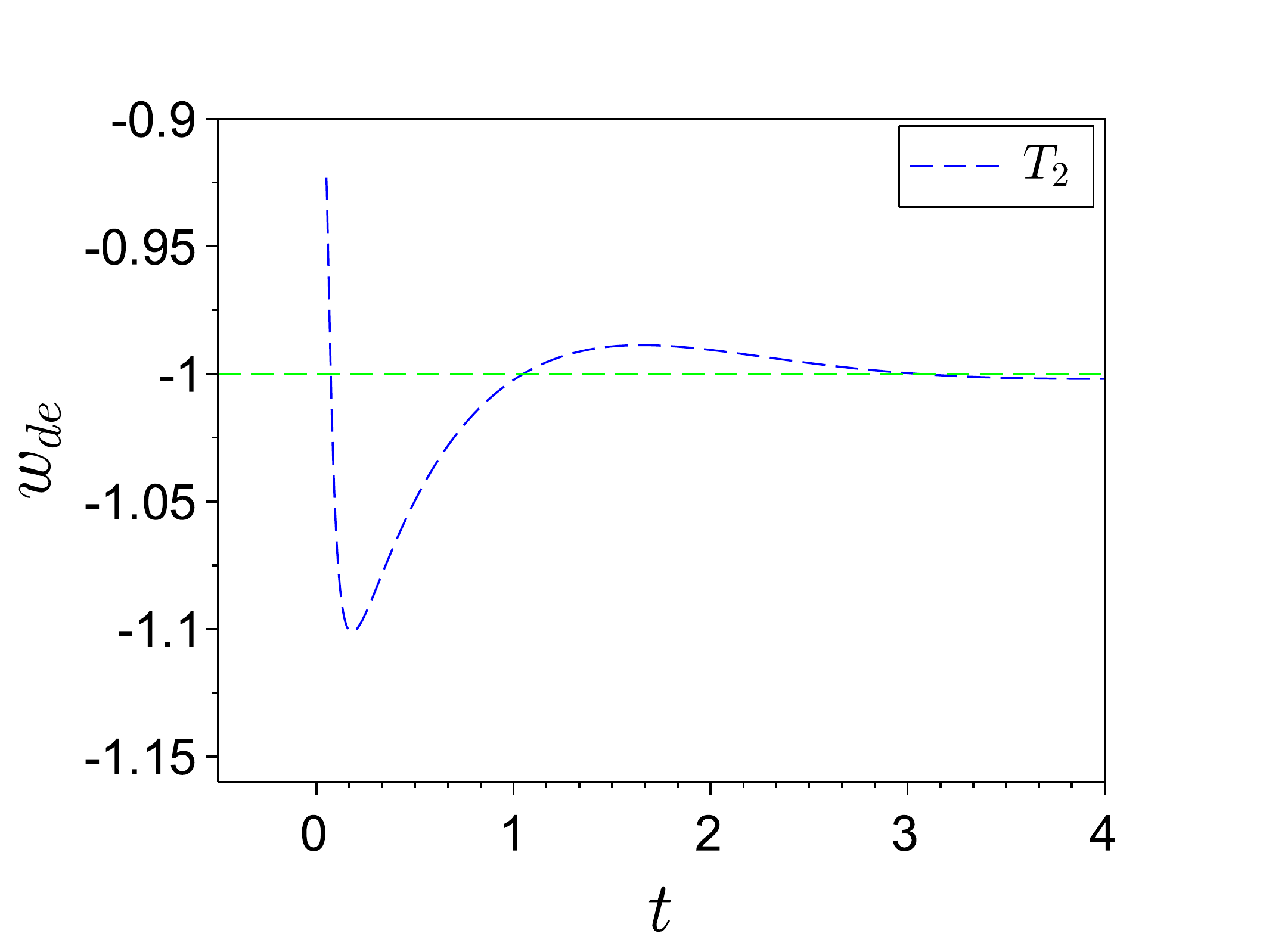}

\medskip

\includegraphics[width=.4\textwidth]{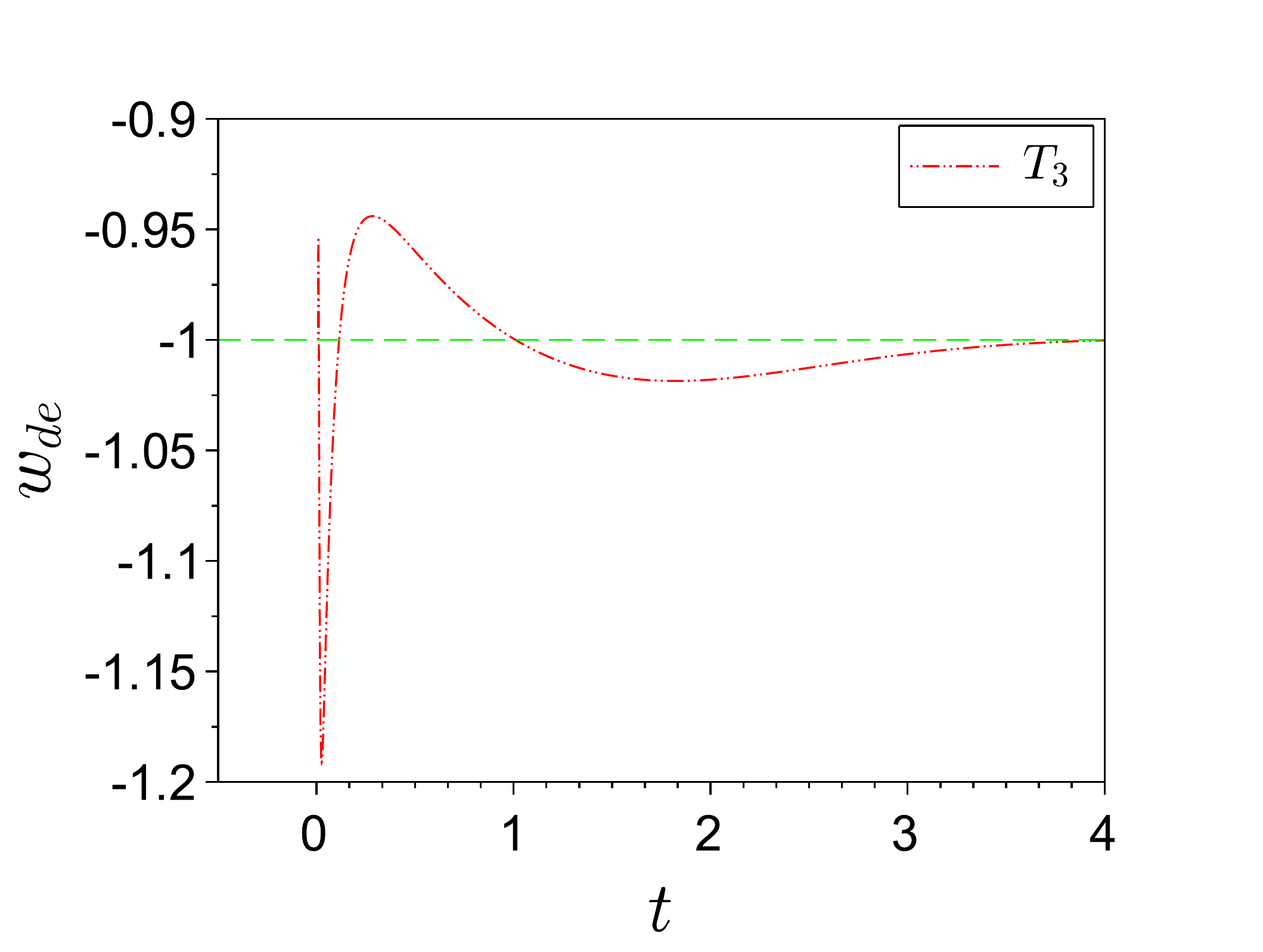}\quad
\includegraphics[width=.4\textwidth]{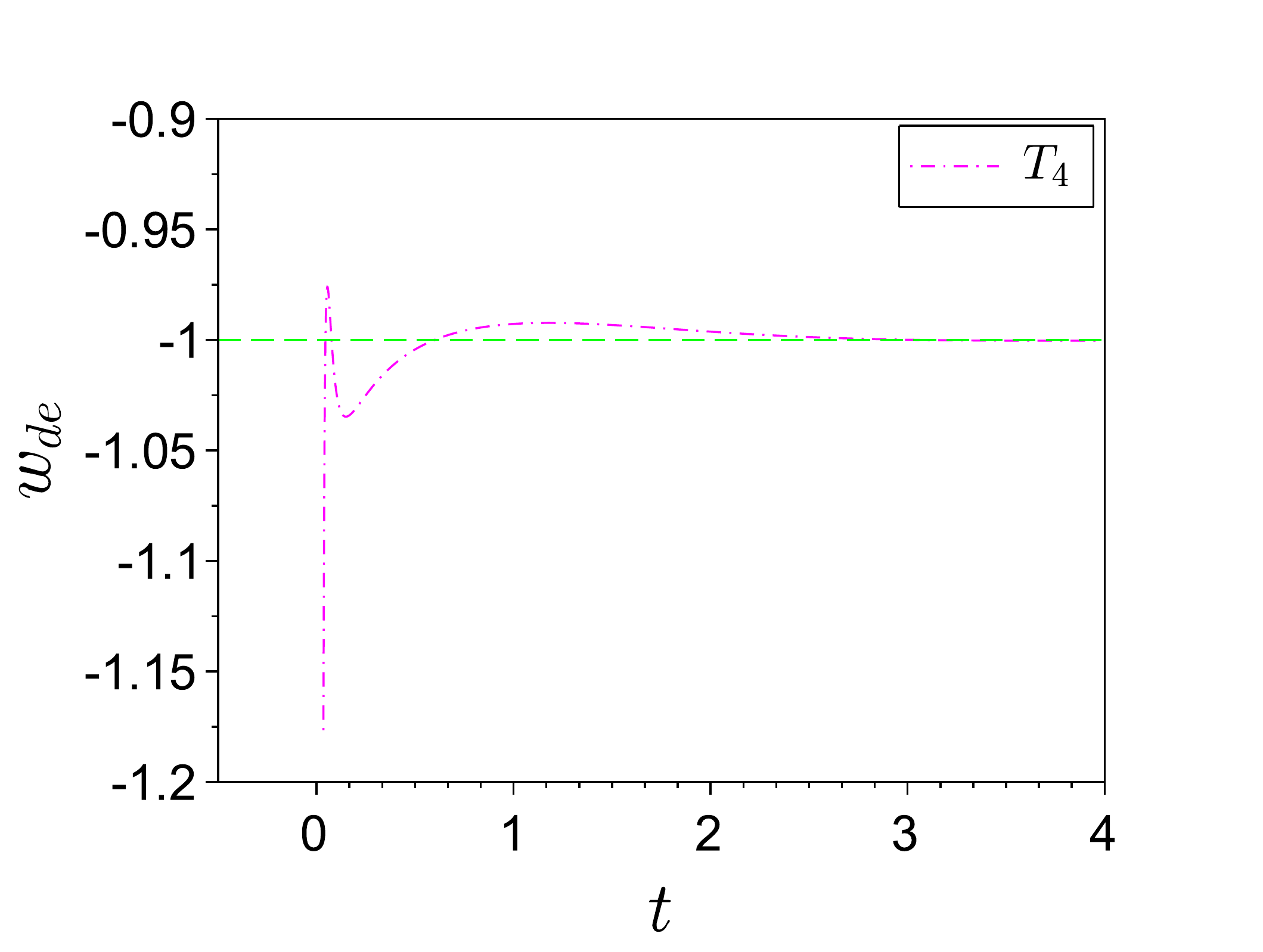}
\caption{The evolution of the dark energy equation of state for scalar torsion coupling models $T_1,T_2,T_3,T_4$}
\label{fig:fig7}
\end{figure}

In Figs.~\ref{fig:fig7}, we show the evolution of the dark energy equation of state in the case of four quintom models non-minimally coupled with the scalar torsion, denoted as $T_i$, with $i=1,..4$, for different coupling parameters and constants. The analysis and the discussion on the numerical
features of the model in the case of scalar torsion coupling models is
presented in the following paragraphs.
\par
A second class of models numerically investigated takes into
account only boundary couplings with the scalar fields, neglecting the scalar torsion couplings
in the quintom model. As previously discussed, we shall also
consider the case of $\xi=-\chi=+1$. In this case, we take the
following models for the coupling functions:
\begin{equation}
f_1(\phi)=0\,,\quad f_2(\sigma)=0\,,\quad g_1(\phi)=c_3\phi^2\,,\quad
g_2(\sigma)=-c_3\sigma^2\,.
\end{equation}
\par
Assuming again that the scalar potential is separable into two exponentials as \eqref{expon}, the energy
densities and pressure for the scalar fields become
\begin{eqnarray}
\rho_{\phi}&=&\frac{1}{2}\dot{\phi}^2+V_1(\phi)+6c_3H\phi\dot{\phi}\,,\\
\rho_{\sigma}&=&-\frac{1}{2}\dot{\sigma}^2+V_2(\sigma)-6c_3H\sigma\dot{\sigma}\,,\\
P_{\phi}&=&\frac{1}{2}\dot{\phi}^2-2c_3(\dot{\phi}^2+\phi\ddot{\phi})-V_1(\phi)\,,\\
P_{\sigma}&=&-\frac{1}{2}\dot{\sigma}^2+2c_3(\dot{\sigma}^2+\sigma\ddot{\sigma})-V_2(\sigma)\,.
\end{eqnarray}
\par
The dark energy equation of state is:
\begin{equation}
w_{\rm de}=\frac{P_{\phi}+P_{\sigma}}{\rho_{\phi}+\rho_{\sigma}}=\frac{\frac{1}{2}\dot{\phi}^2-2c_3(\dot{\phi}^2+\phi\ddot{\phi})-V_1(\phi)-\frac{1}{2}\dot{\sigma}^2+2c_3(\dot{\sigma}^2+\sigma\ddot{\sigma})-V_2(\sigma)}{\frac{1}{2}\dot{\phi}^2+V_1(\phi)+6c_3H\phi\dot{\phi}-\frac{1}{2}\dot{\sigma}^2+V_2(\sigma)-6c_3H\sigma\dot{\sigma}}\,.
\end{equation}
\par
For boundary coupling with the scalar fields, the Klein-Gordon equations take the form:
\begin{eqnarray}
\label{klein11}
\ddot{\phi}+3H\dot{\phi}+6c_3\phi(3H^2+\dot{H})+\frac{dV_1(\phi)}{d\phi}&=&0\,,\\
\label{klein22}
\ddot{\sigma}+3H\dot{\sigma}+6c_3\sigma(3H^2+\dot{H})-\frac{dV_2(\sigma)}{d\sigma}&=&0\,.
\end{eqnarray}
\par
As in the previous case, we express the acceleration equation for
the scale factor as
\cite{num1,num2,marciu1}:
\begin{equation}
\label{accelerationeqs2}
\frac{\ddot{a}}{a}=-\frac{1}{6}(\rho_{\phi}+\rho_{\sigma}+3
P_{\phi}+ 3P_{\sigma} )-\frac{\Omega_{\rm m 0}H_{0}^2}{2 a^3},
\end{equation}
and we study the system of equations numerically in a similar way
to the scalar torsion coupling with the scalar fields. For these cases, we
present the dynamics of four different models, denoted
$B_i$ with $i=1,..,4$, taking different parameters and
constants.
\par
\begin{figure}[H]
\centering
\includegraphics[width=.4\textwidth]{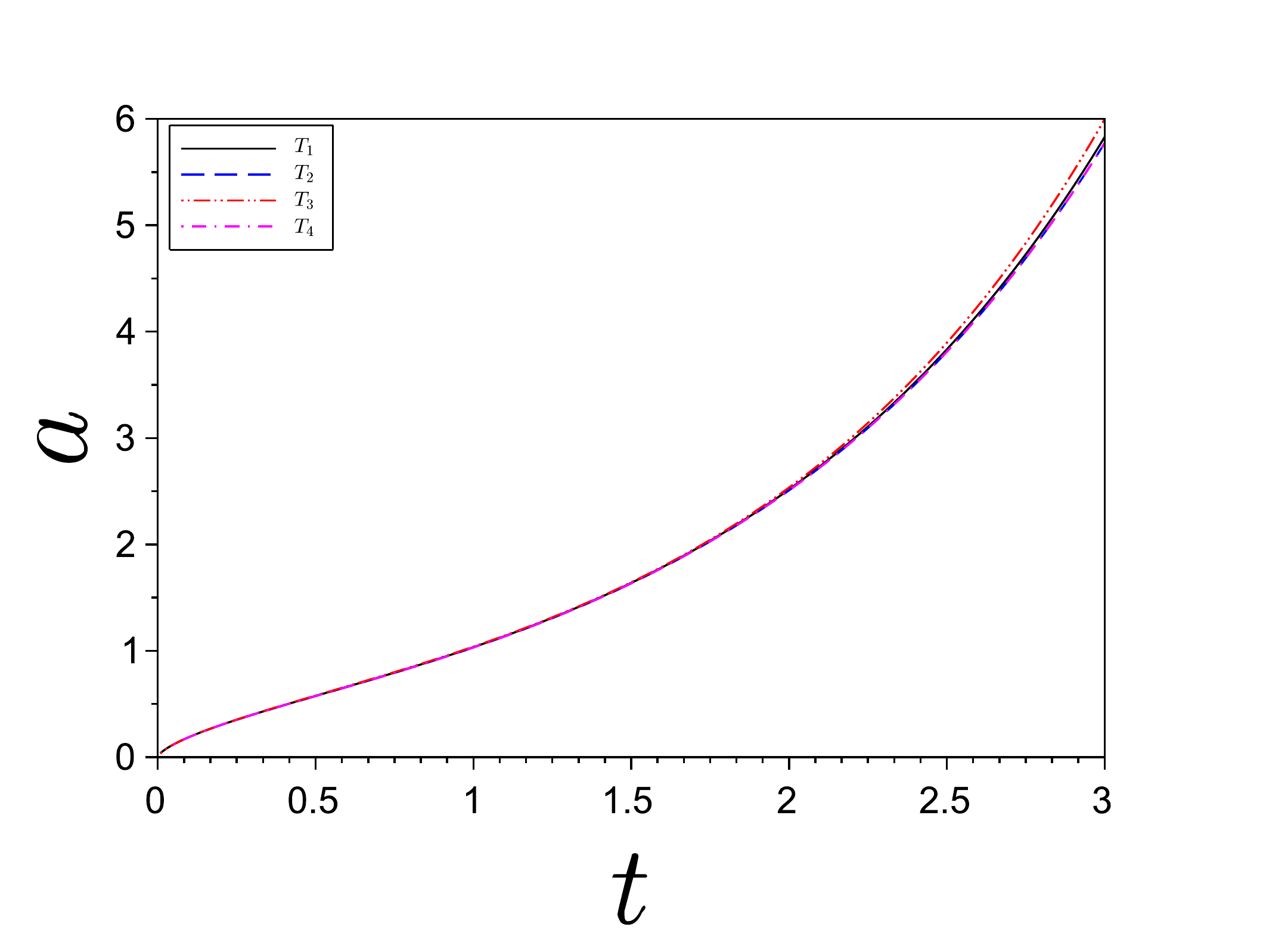}\quad
\includegraphics[width=.4\textwidth]{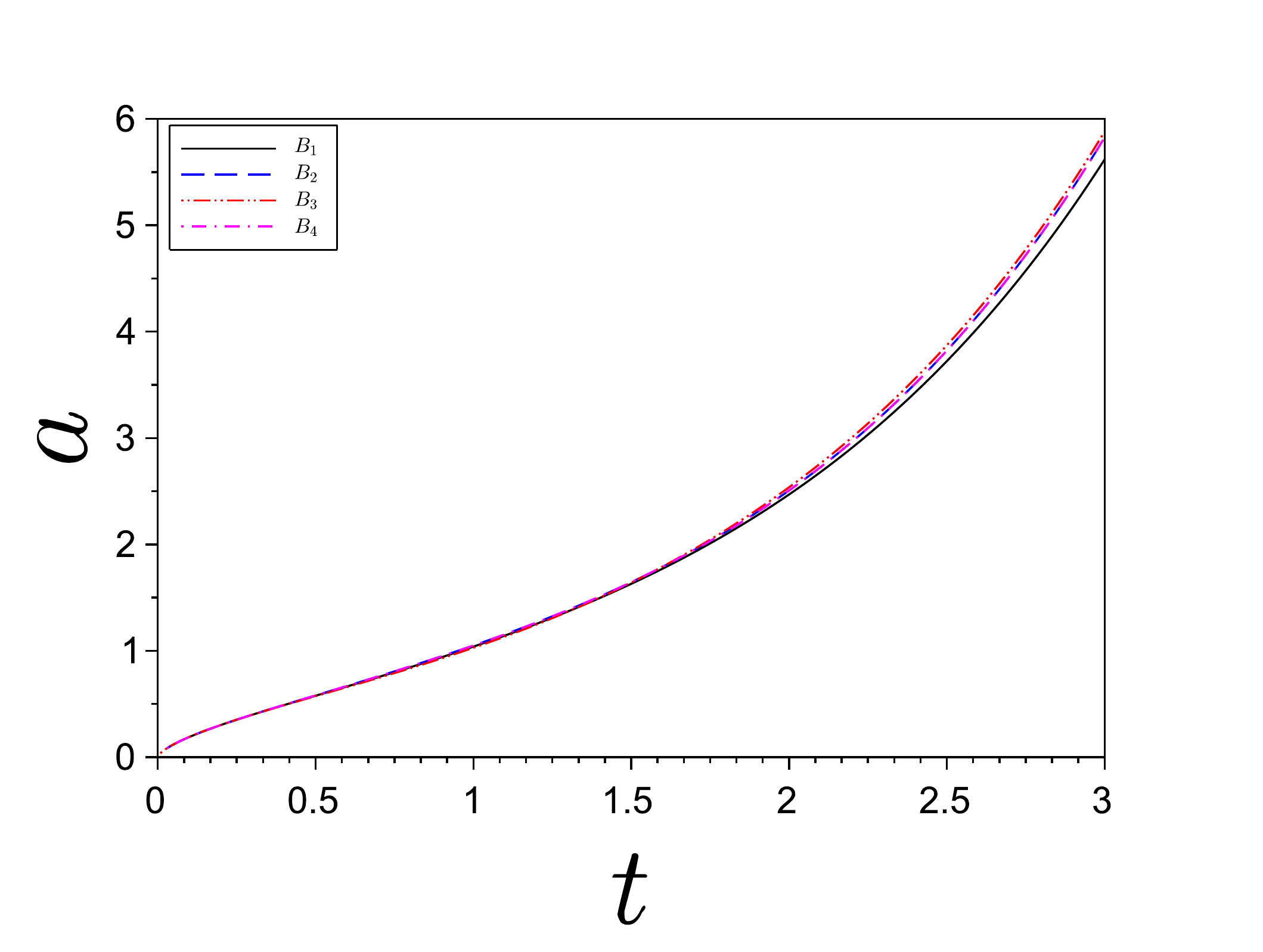}\quad

\caption{The time evolution of the cosmic scale factor for models with scalar torsion coupling(left panel) $T_1,T_2,T_3,T_4$ and boundary couplings(right panel) $B_1,B_2,B_3,B_4$}
\label{fig:fig1}
\end{figure}
\begin{figure}[H]
\centering
\includegraphics[width=.4\textwidth]{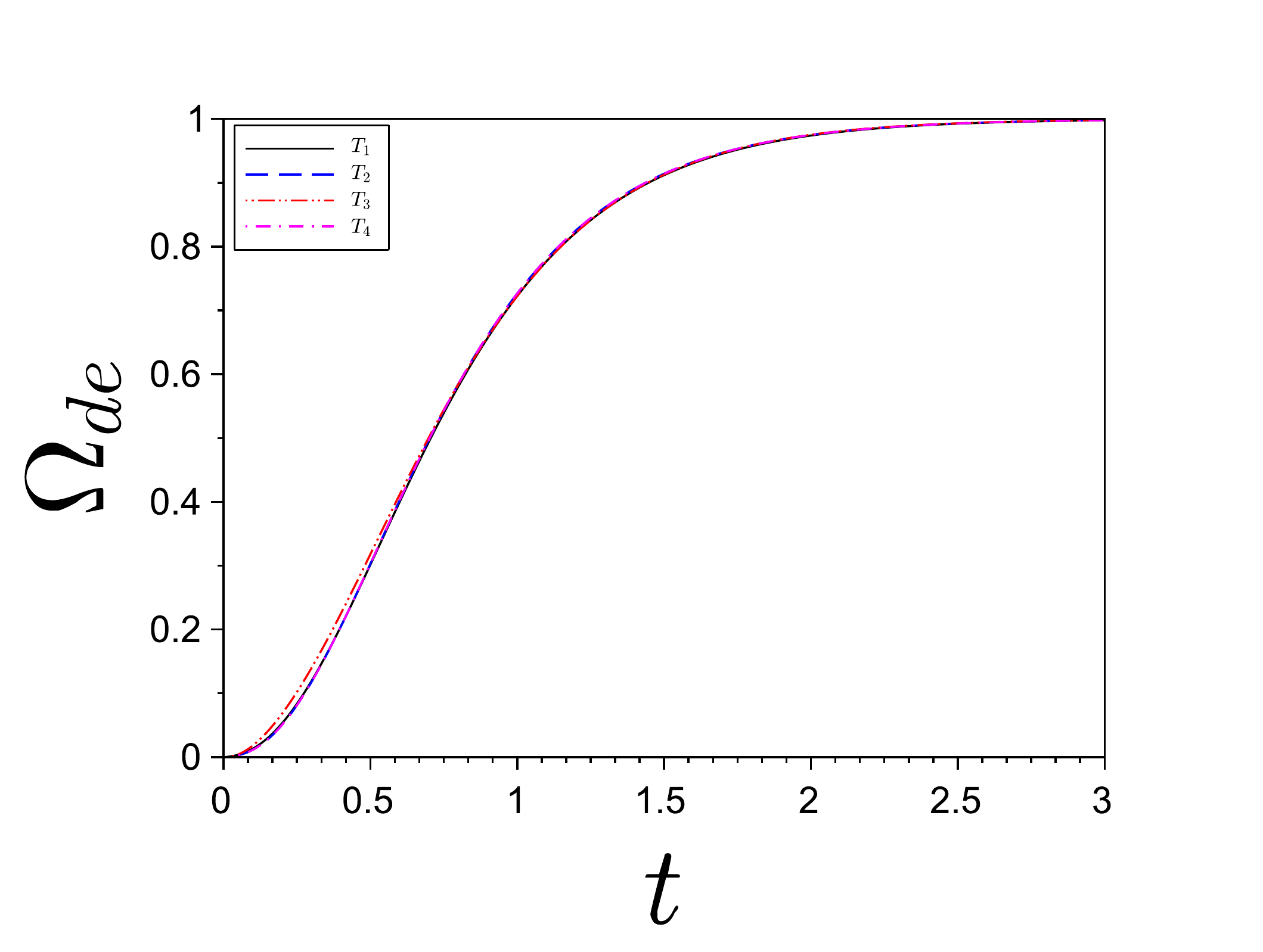}\quad
\includegraphics[width=.4\textwidth]{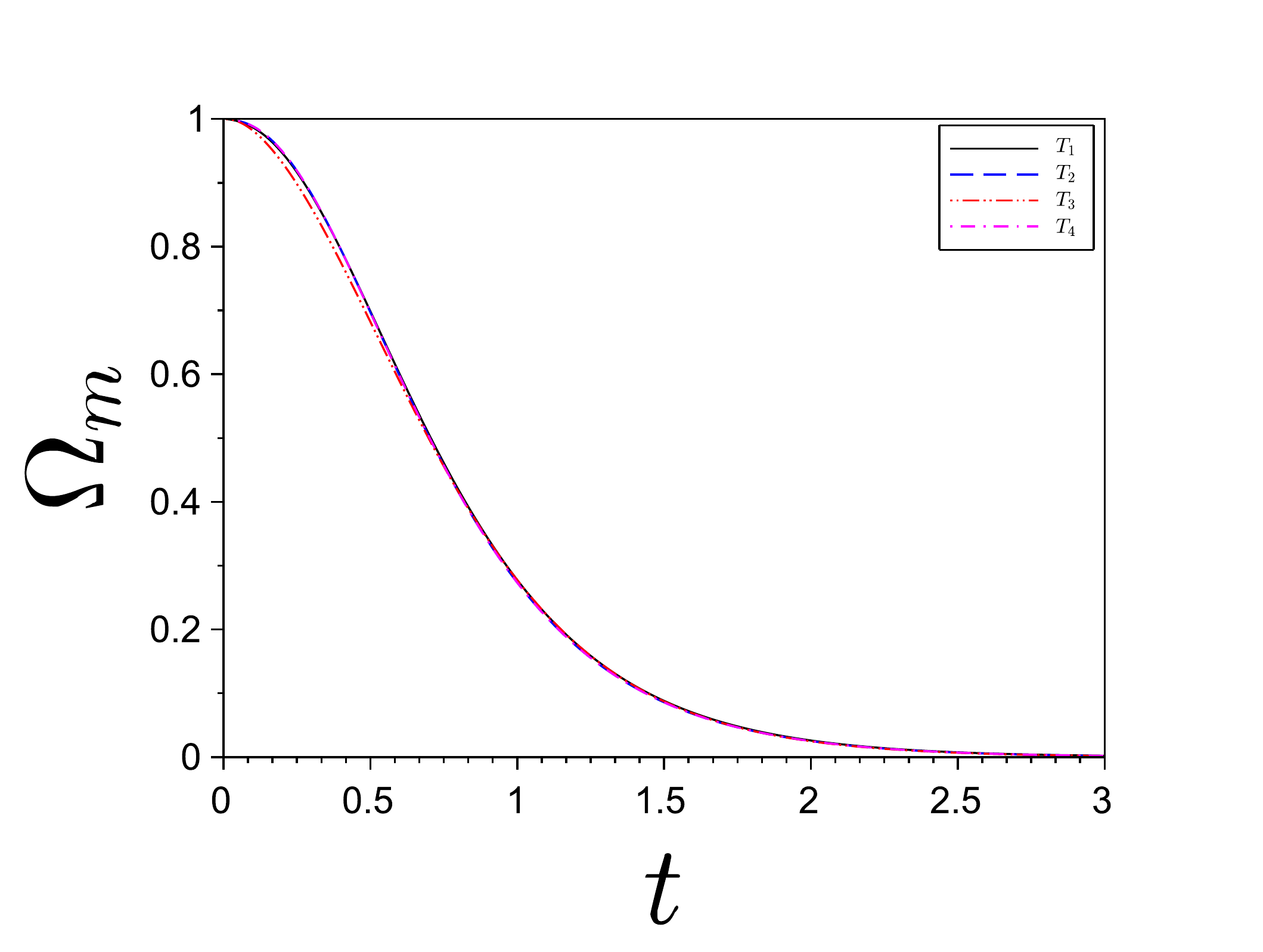}\quad

\caption{The evolution of the quintom energy density and matter energy density for scalar torsion coupling models}
\label{fig:fig3}
\end{figure}
\begin{figure}[H]
\centering
\includegraphics[width=.4\textwidth]{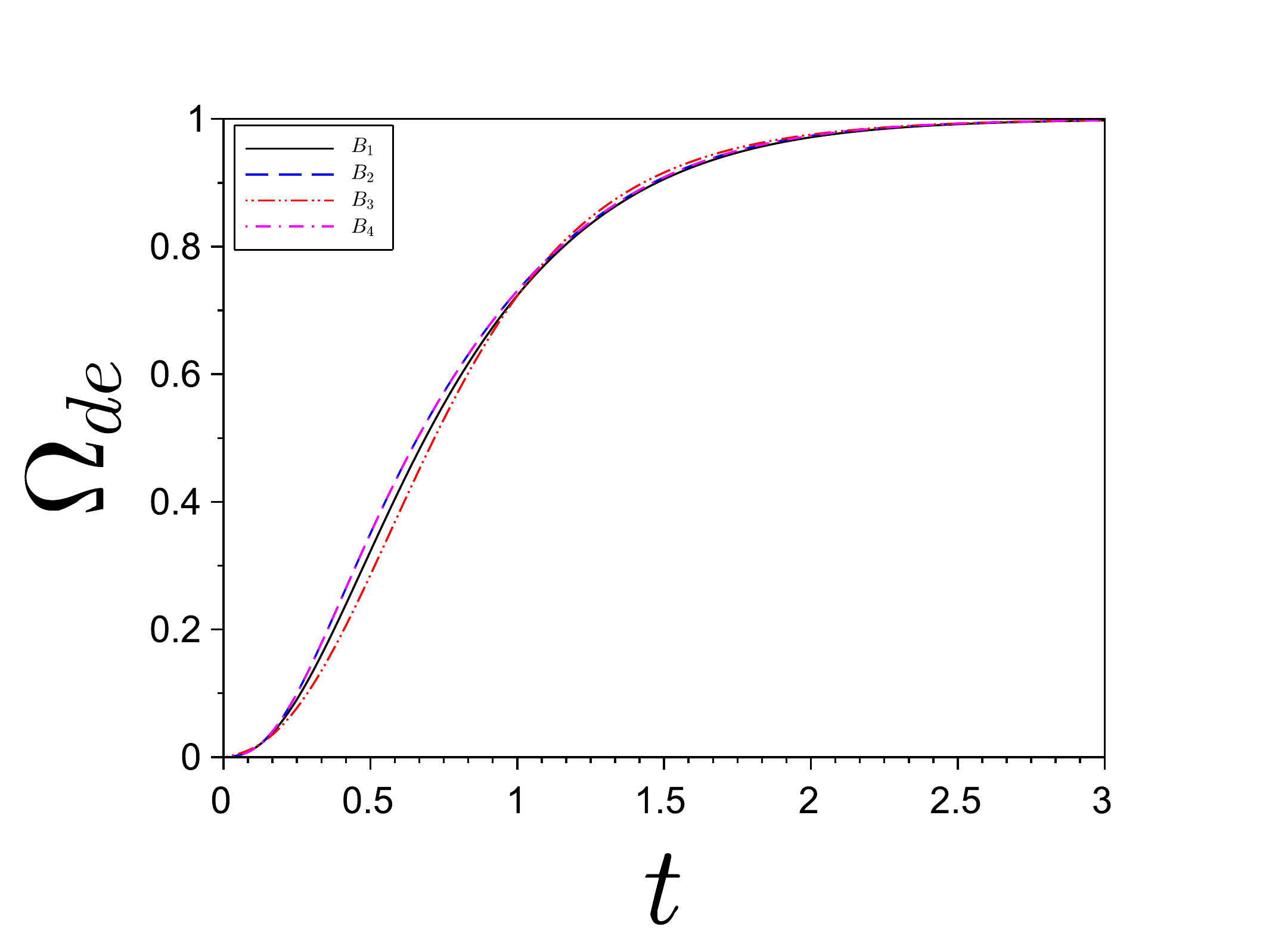}\quad
\includegraphics[width=.4\textwidth]{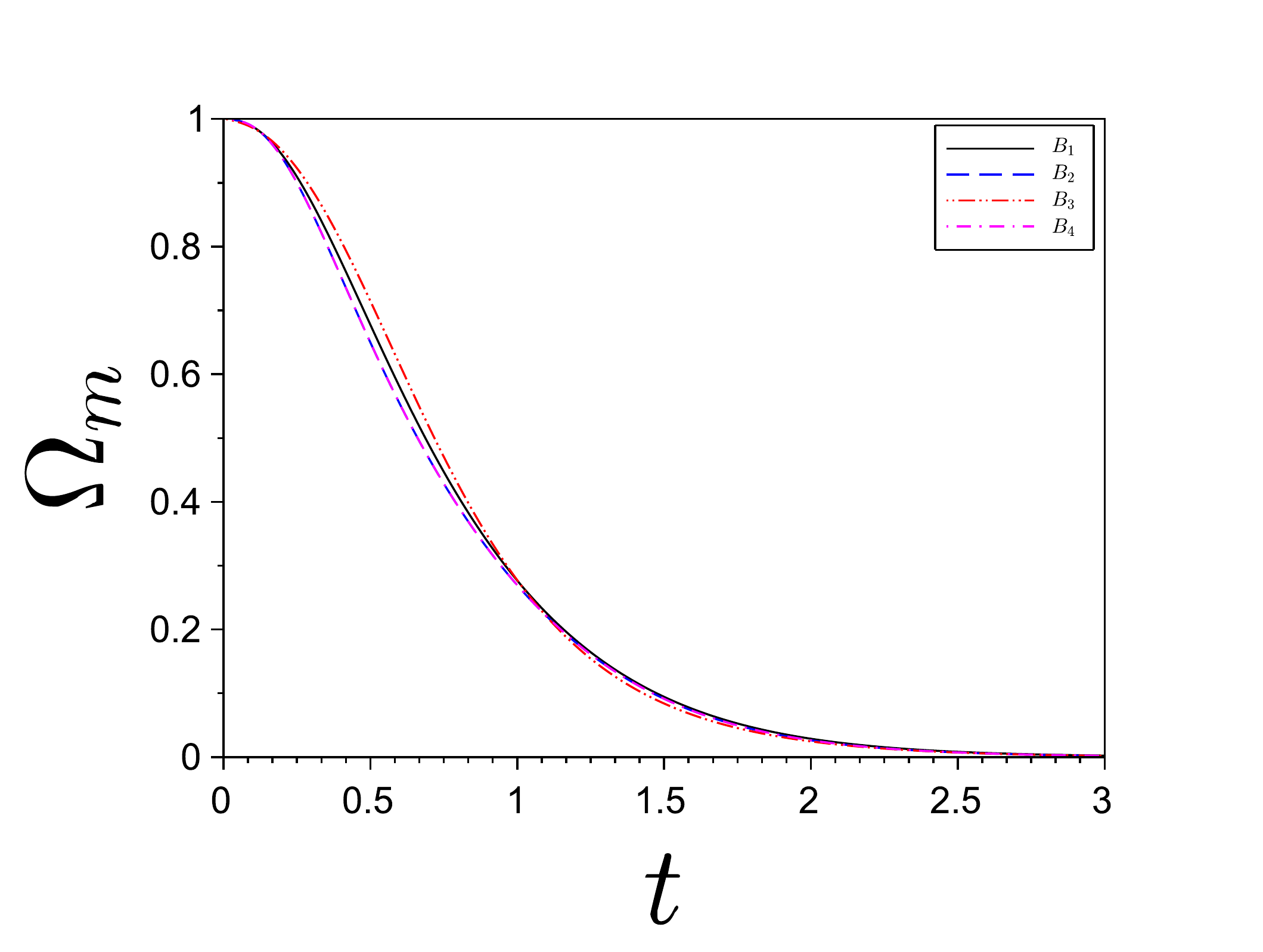}\quad

\caption{The evolution of quintom energy density and matter energy density for boundary coupling models}
\label{fig:fig5}
\end{figure}
\begin{figure}[H]
\centering
\includegraphics[width=.4\textwidth]{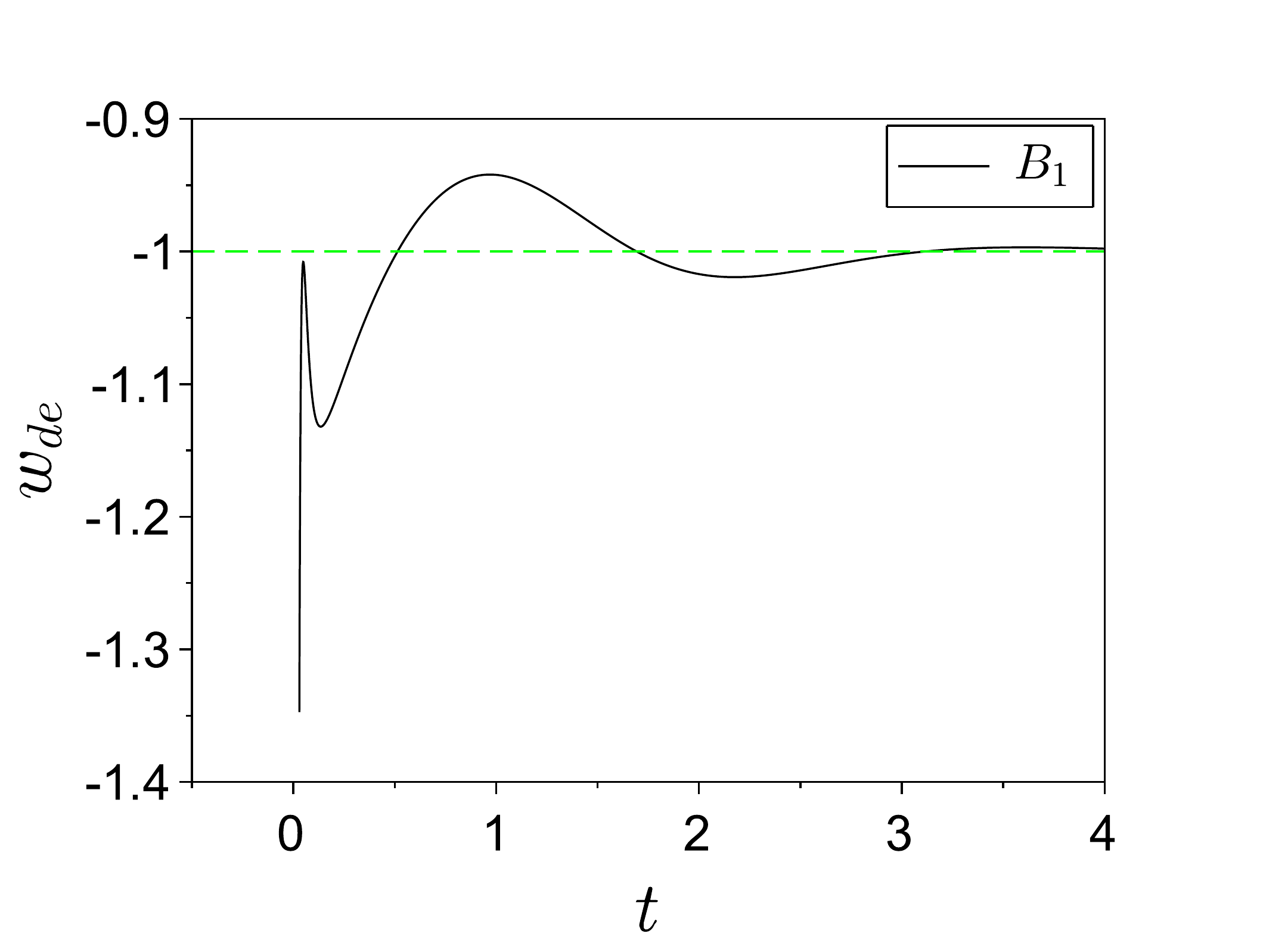}\quad
\includegraphics[width=.4\textwidth]{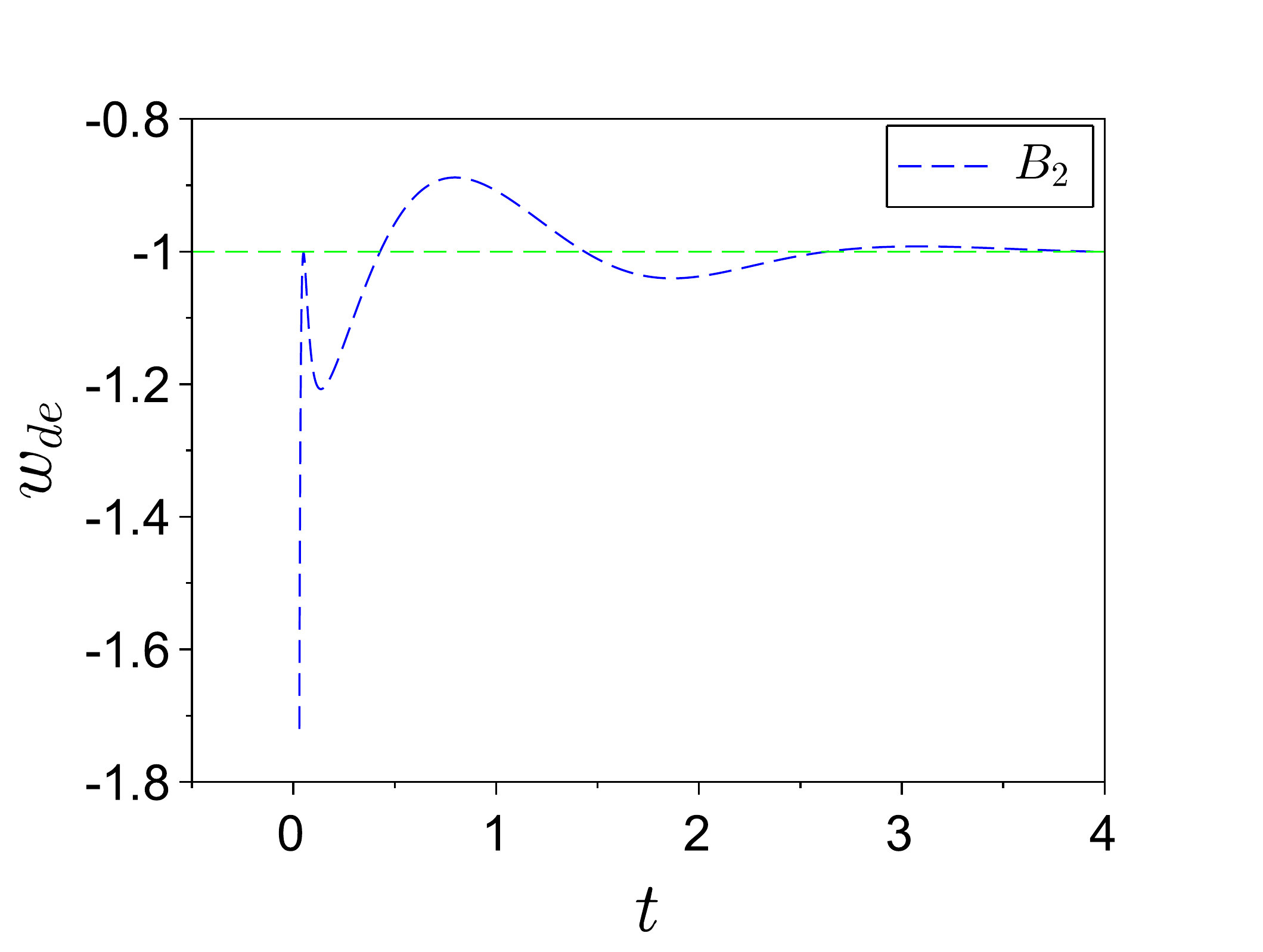}

\medskip

\includegraphics[width=.4\textwidth]{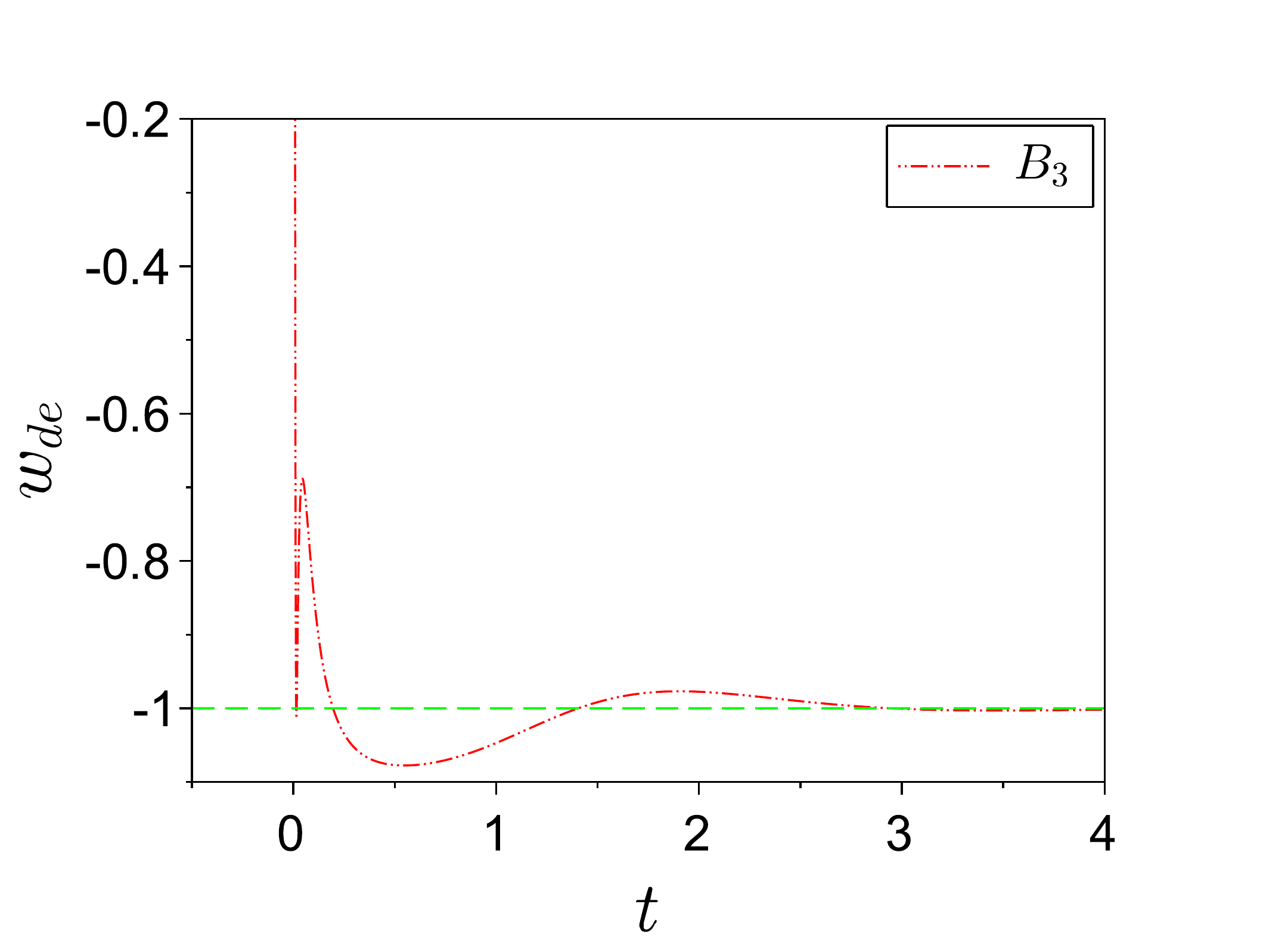}\quad
\includegraphics[width=.4\textwidth]{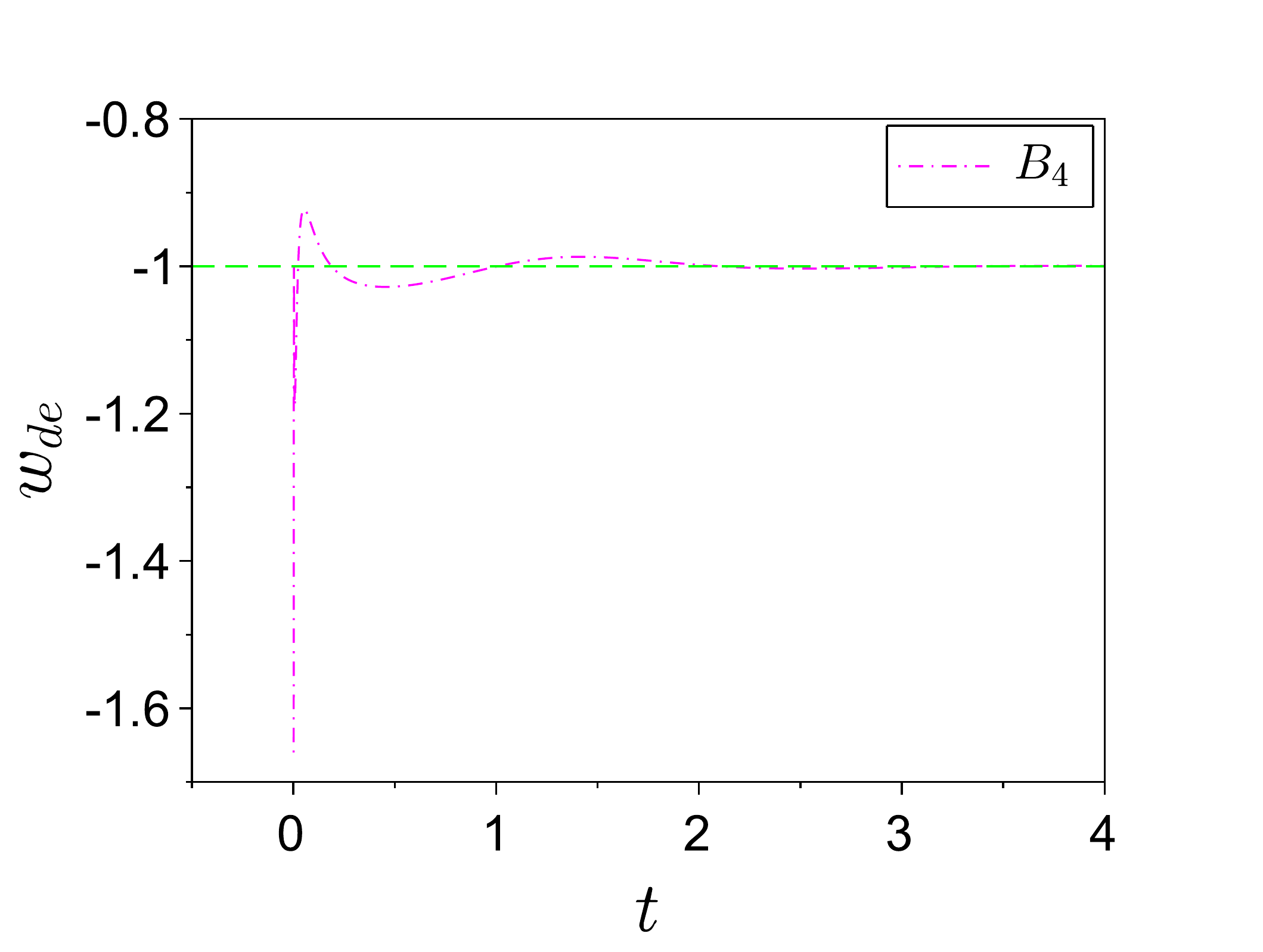}

\caption{The evolution of the dark energy equation of state for boundary coupling models $B_1,B_2,B_3,B_4$}
\label{fig:fig11}
\end{figure}

In the following, we shall analyse the dynamical properties of the
quintom model by investigating the numerical results obtained. As
previously stated, the numerical solutions deduced in the case of
scalar torsion and boundary couplings with the scalar fields are depicted in Figs.~\ref{fig:fig1}-\ref{fig:fig11}. As a general remark, we
observe that these results are compatible to the previous ones
obtained in the case of a quintom model non-minimally coupled
with scalar curvature \cite{marciu1}. Then,
we shall discuss the obtained results and the possible physical
consequences of the scalar torsion and boundary couplings with the scalar fields in the quintom
model.
\par
In the left panel of Fig.~\ref{fig:fig1}, we have represented the evolution of
the cosmic scale factor as a function of cosmic time for the four scalar
torsion models previously mentioned $T_i$, for
different initial conditions. For the scalar torsion model $T_1$, the
initial conditions are: $c_1=0.7,
\lambda_{1}=\lambda_{2}=0.6,  V_{1}=V_{2}=1.02,
\phi(t_i)=\sigma(t_i)=-0.6, \dot{\phi}(t_i)=\dot{\sigma}(t_i)=0,
t_i=0.008$. The second scalar torsion model, denoted as $T_2$, has the
following initial conditions: $c_1=0.5,
\lambda_{1}=\lambda_{2}=0.5,  V_{1}=V_{2}=1,
\phi(t_i)=\sigma(t_i)=0.5, \dot{\phi}(t_i)=0.01,
\dot{\sigma}(t_i)=0.001, t_i=0.05$. In the case of the third
scalar torsion model $T_3$, we have the initial conditions: $c_1=0.7,
\lambda_{1}=\lambda_{2}=0.6,  V_{1}=V_{2}=1.08,
\phi(t_i)=\sigma(t_i)=0.6, \dot{\phi}(t_i)=0, \dot{\sigma}(t_i)=0,
t_i=0.008$. The last scalar torsion model, named $T_4$, has the
conditions as: $c_1=0.6, \lambda_{1}=\lambda_{2}=0.01,
V_{1}=V_{2}=1.01, \phi(t_i)=\sigma(t_i)=8,
\dot{\phi}(t_i)=0.0001, \dot{\sigma}(t_i)=0.001, t_i=0.035$. We
can observe that the dynamics of the Universe in the case of the
four scalar torsion models corresponds to an accelerated expansion, very close to a de Sitter expansion at late times. The present time is at the numerical
value of $t_0=0.96$, where the cosmic scale factor is
approximately equal to unity $a(t_0)\sim 1$, as requested from the
numerical method considered. Moreover, we can observe that the
values of the coupling coefficient $c_1$ have minor influence on
the dynamics at the large scale. The Universe in this model is accelerating independently from the values of the
coupling parameter $c_1$.
\par
The case of boundary coupling models, neglecting the scalar torsion
couplings are represented in the right panel of Fig.~\ref{fig:fig1},
where we observe a similar behavior - the dynamics of the
Universe in the case of the four boundary coupling models
corresponds to an accelerated expansion, toward a de Sitter stage, independently from the values of the coupling
parameter $c_3$. The boundary coupling models investigated in this
figure are denoted as $B_i$, and we
shall describe the initial conditions used in order to investigate
the effects of the boundary couplings. In the case of the boundary
coupling model $B_1$, we have used the following initial
conditions: $c_3=0.5, \lambda_{1}=\lambda_{2}=0.7,
V_{1}=V_{2}=1, \phi(t_i)=\sigma(t_i)=0.6,
\dot{\phi}(t_i)=0.001, \dot{\sigma}(t_i)=0.000001, t_i=0.0295$.
The next model $B_2$ has the initial conditions: $c_3=0.6,
\lambda_{1}=\lambda_{2}=0.4,  V_{1}=V_{2}=1.05,
\phi(t_i)=\sigma(t_i)=1.5, \dot{\phi}(t_i)=0.001,
\dot{\sigma}(t_i)=0.001, t_i=0.0295$. The third boundary coupling
model, $B_3$ has: $c_3=0.4, \lambda_{1}=\lambda_{2}=0.2,
V_{1}=V_{1}=1.05, \phi(t_i)=\sigma(t_i)=-5,
\dot{\phi}(t_i)=0.00001, \dot{\sigma}(t_i)=0.00001, t_i=0.0083$.
Finally, for the last boundary coupling model $B_4$, we have
considered: $c_3=0.6, \lambda_{1}=\lambda_{2}=0.5,
V_{1}=V_{2}=1.055, \phi(t_i)=\sigma(t_i)=1.1,
\dot{\phi}(t_i)=\dot{\sigma}(t_i)=0.00001, t_i=0.0015$. As
previously mentioned, we have a similar behavior as in the left panel of 
Fig.~\ref{fig:fig1}, an evolution of the system toward a de Sitter stage in the distant future.
\par
The influence of the scalar torsion coupling coefficient $c_1$ in the
evolution of the density parameters of the Universe, dark energy
density parameter and matter density parameters, respectively, are
presented in Fig.~\ref{fig:fig3}. In these
figures we can observe that at the present time $t_0=0.96$, the
density parameters have approximately the values suggested by
astrophysical observations, $\Omega_{\rm de}=1-\Omega_{\rm m} \sim 0.70$.
From the numerical evolution, it can be noticed that at the
initial time, the Universe is deep in the matter dominated era,
while at the final numerical time, the cosmic picture is dominated
by the dark energy quintom fields non-minimally coupled with
the scalar torsion. Hence, the present quintom model with a scalar torsion coupling is
able to explain the current values of the density parameters in
the Universe, in a good agreement with astronomical and
astrophysical observations. It is easy to see that the values of the
coupling coefficient $c_1$ have minor effects on the values of the
density parameters, since with a proper fine-tuning of the initial
conditions, we are able to reproduce the corresponding current
values of the density parameters in the Universe. From this, we
remark that the present generalised quintom model in the framework of 
teleparallel gravity theory with a scalar torsion coupling is a
feasible dark energy model.
\par
Next, in Fig.~\ref{fig:fig5} is displayed the
evolution of the density parameters in the case of boundary
coupling models $B_i$, considering different values of
the coupling coefficient $c_3$, previously discussed. Here, we have a
similar behavior as in the case of a scalar torsion coupling. At the
initial time, the Universe is deep in the matter epoch, while at
late times the dark energy fields dominate the cosmic picture.
Hence, boundary coupling models $B_i$ are representing
also feasible dark energy models, explaining the acceleration of the
Universe as well as the current values of the density parameters.
The values of the boundary coupling parameters $c_3$ have minor
influence on the evolution of the density parameters corresponding
to the dark energy fluid and matter fluid, respectively, similar
to the scalar torsion models $T_i$. Consequently, in both models
with boundary couplings $B_i$ and scalar torsion couplings $T_i$ endowed
with decomposable exponential potentials, the Universe is evolving towards a state dominated by dark energy
fields over the matter fluid, while the cosmic scale factor is in an accelerated stage. 
\par
The time evolution of the dark energy equation of state in the
case of scalar torsion coupling models $T_i$ is shown
in Fig.~\ref{fig:fig7}. We can remark that for
the scalar torsion coupling models, the dark energy equation of state can
exhibit the main specific feature to quintom models; the crossing
of the phantom divide line. For scalar torsion coupling models,
in the first stage of evolution, the dark energy equation of state
presents oscillations around the cosmological constant boundary,
while at later times the equation of state evolves asymptotically
towards the $\Lambda$CDM model, acting almost as the cosmological
constant. As a consequence, the generalised quintom model in the
teleparallel gravity theory with a scalar torsion coupling can be in
agreement with cosmological observations which have suggested
that the cosmological constant boundary might be crossed by the
dark energy equation of state.
\par
Finally, the evolution of the dark energy equation of state in the
case of boundary coupling models $B_i$, is
depicted in Fig.~\ref{fig:fig11}. As in the
previous models, at the initial stage of evolution, the models show
the crossing of the phantom divide line, while in the end
at the final time, the dark energy quintom model acts asymptotically
as a cosmological constant. By comparing the evolution of the
boundary coupling models $B_i$ with the scalar torsion models
$T_i$, one can notice that in the case of boundary
coupling, the dark energy equation of state presents pronounced
oscillations around the phantom divide line. Hence, for scalar
torsion models $T_i$, the oscillations around the
phantom divide line are more reduced. As a concluding remark, we
remind that the present generalised quintom model in the
teleparallel gravity with scalar torsion and boundary
couplings represents also a possible dark energy model, an alternative to
the $\Lambda$CDM model, which can explain the observed crossing of
the cosmological constant boundary in the near past of the dark
energy equation of state, as suggested by various cosmological
observations. Notice that this behaviour can be achieved without evoking any cosmological constant.

\section{Conclusions}
In this paper we have considered a general quintom model constructed in the teleparallel framework, taking into account a possible non-minimal interaction between the quintom fields and the scalar torsion $T$ and boundary terms $B$, respectively. Our Lagrangian is constructed in the teleparallel alternative of general relativity framework, where the boundary term $B$ present in the action of the model is related to the divergence of the torsion vector. We have analysed the structure of the phase space, taking into account that the potential of our model is decomposed into a sum of two exponential terms, revealing the effects of the non-minimal coupling of the scalar torsion and boundary terms for the dynamics of our quintom model. We should note that the choice of the potential for the dark energy fields play an important role in the phase space dynamics, the results of our analysis are specific for this type of potential. By studying the critical points of the autonomous system, we have revealed the effects of the non-minimal coupling coefficients for the main characteristics of our critical points, related to the position in the 6D dimensional space and the corresponding eigenvalues, which determine the stability of the dynamical solutions. As can be noticed from the analysis, the matter dominated epoch is characterised by a critical point with a saddle behavior. Our analysis revealed that  the remaining critical points of the autonomous dynamical system are characterizing a dark energy dominated universe. Although the complexity of our model is high due to the presence of the $6$ constant parameters related to the potential energy, the couplings of the scalar fields with the scalar torsion and boundary terms, respectively, we have obtained possible constraints for the parameters of our model which correspond to different dynamical scenarios. After the dynamical study of the present quintom scenario, we have considered a numerical evolution of the system of equations from the matter dominated era, considering particular choices for the  scalar torsion and boundary coupling terms. The numerical study takes into account that in the matter dominated epoch the cosmic scale factor satisfy the usual dynamical behavior and the fields are almost frozen. The analysis revealed that the system evolves towards to a de Sitter epoch in the distant future. As expected, for different values of the scalar torsion and boundary coupling parameters, the equation of state for the dark energy component presents oscillations around the $\Lambda$CDM behavior, and the oscillatory characteristic of the model is present for the scalar torsion and boundary couplings, respectively. Although the late time dynamics of our model corresponds to a $\Lambda$CDM evolution, in the early times the dark energy equation of state presents a quintom behaviour in favor with the recent cosmological observations regarding the crossing of the phantom divide line boundary by the dark energy equation of state. Consequently, the present quintom model generalised to the teleparallel equivalent of general relativity represents a feasible cosmological scenario which can explain the large scale behavior of the universe at the level of background dynamics.      

\vspace{.5cm}
\begin{flushleft}
	\textbf{Acknowledgments}
\end{flushleft}
S.B. is supported by the Comisi{\'o}n Nacional de
Investigaci{\'o}n Cient{\'{\i}}fica y Tecnol{\'o}gica (Becas Chile
Grant No.~72150066). S.B. thanks the Yukawa Institute for Theoretical Physics at Kyoto University,
where this work was completed during the workshop YITP-T-17-02 ``Gravity and Cosmology 2018". P.R. acknowledges University Grants Commission, Govt. of India for providing research project grant (No. F.PSW-061/15-16 (ERO)). P.R. also acknowledges Inter
University Centre for Astronomy and Astrophysics (IUCAA), Pune, India, for awarding Visiting Associateship. During this work, During this work, M. Marciu was partially supported by the project 29/2016 ELI-RO from the Institute of Atomic Physics, Magurele. The authors would like to thank C. M. for carefully checking English and typos in the manuscript.\\

\end{document}